%% file: ebls_arxiv_version.tex
\documentclass[pra,aps,twocolumn,groupedaddress,nofootinbib]{revtex4}
\input{ebls_preamble.tex}

\usepackage{dcolumn}
\newcommand{\mb}{\mathsf{MaxBody}} 
\newcommand{\mr}{\mathsf{MaxRange}} 

\begin{document}

\title{RG and Stability in the Exciton Bose Liquid}

\author{Ethan Lake}
\email{elake@mit.edu}
\affiliation{Department of Physics, Massachusetts Institute of Technology, Cambridge, MA, 02139}

\begin{abstract}
	The exciton Bose liquid (EBL) is a hypothesized phase of bosons in 2+1D which possesses a dispersion that is gapless along the coordinate axes in momentum space. The low energy theory of the EBL involves modes on all length scales, extending all the way down to the lattice spacing. In this note, we discuss an RG scheme that can be used to address the stability of this and related phases of matter. We find that in the absence of an extensively large symmetry group, realizing the simplest formulation of the EBL always requires fine-tuning. However, we also argue that the addition of certain marginal interactions can be used to realize a stable phase, without the need for fine-tuning. A simple generalization to 3+1D is also discussed. 
		
\end{abstract}

\maketitle

\section{Introduction} 

Most fixed points of renormalization group (RG) flows are characterized by the absence of any intrinsic length scale. At such fixed points the low energy physics is scale-invariant (and often conformally invariant), with all the universal data of the fixed point being determined by pure dimensionless numbers.
% Often scale invariance is further enlarged to conformal invariance, and techniques from CFT can be brought to bear. 

One important counterexample is a Fermi liquid. Fermi liquids can be realized as gapless endpoints of RG flows, but they are not scale-invariant: the Fermi momentum $k_F$ plays a crucial role in determining correlation functions, and low energy degrees of freedom live at length scales all the way down to $ 1/k_F$, which is usually comparable to the lattice spacing. 

Another less well-known example of a fixed point with an intrinsic length scale is the exciton Bose liquid (EBL), which was proposed in \cite{paramekanti2002ring} and which has recently arisen in a diverse array of different contexts (e.g. \cite{you2020fracton,you2020emergent,gorantla2021low,seiberg2003exotic}). The EBL is a phase of bosons in 2+1D possessing a dispersion of the form 
\be \label{intro_disp} \ep_\bfk \propto |\sin(k_xa/2)\sin(k_ya/2)|,\ee 
where $a$ is the lattice spacing. The most important aspect of \eqref{intro_disp} is that $\ep_\bfk = 0$ along the coordinate axes in momentum space. This means that the theory possesses low energy modes on all length scales down to the lattice spacing, and consequently exhibits ``UV-IR mixing'' \cite{gorantla2021low,seiberg2003exotic}. As such, any theory capturing the universal physics of this phase of matter cannot be fully scale-invariant, as it must know about the scale $a$.

The EBL was originally proposed \cite{paramekanti2002ring} to be a stable phase of matter, viz. one which does not require fine-tuning to be realized. This claim was however later disputed \cite{tay2011possible}. In order to have a clear answer to the question of whether or not the EBL is stable (in the presence of a given group of microscopic symmetries), one needs to construct an RG scheme that determines which perturbations affect the universal physics. As far as the author is aware, there does not seem to be a discussion of how to do this in the literature (at least when the thermodynamic limit is taken), and in this note we will attempt to fill in this gap. 

In order to address the question of stability, we need to understand how to perform RG in a system where low energy modes live at all length scales. 
RG is often described as a procedure involving coarse-graining in space, a la Kadanoff's original spin-blocking procedure \cite{kadanoff1966scaling}. There is however no unique way of performing RG, and the most useful scheme will depend on context.

In general, a useful RG scheme is one which eliminates non-universal degrees of freedom, namely those which are not necessary for describing the low energy physics of the system. 
These non-universal degrees of freedom may or may not be associated with short distance scales. 
 In particular, associating RG flow with a successive elimination of short distance degrees of freedom is only appropriate in problems where things happening at short distances also live at high energies, which is not the case for the EBL.
% , viz. when there is a separation of scales. 
It is therefore misleading to describe the EBL and related models as being ``beyond renormalization'' \cite{you2021fractonic,you2021bfractonic}\footnote{One would certainly not use these words when discussing a Fermi liquid, for example.}; rather, 
such models simply mandate an RG scheme which does not proceed by eliminating short-distance modes. The purpose of this note is to construct an appropriate RG scheme, and determine whether or not the EBL is in fact stable.

An outline of the remainder of this note is as follows. In section \ref{sec:rev}, we give a brief review of the aspects of the EBL which will be relevant in the following sections. Section \ref{sec:rg} describes our approach to performing RG, and in \ref{sec:stab} we use this approach to analyze the stability of the EBL. We will find that the simplest version of the EBL can only be realized with fine-tuning, but that a certain choice of marginal ``Landau parameters'' can be made such that a stable phase seems likely to exist. In section \ref{sec:3d} we briefly describe a generalization to a related 3+1D model, and we conclude in section \ref{sec:conc}.

\section{The EBL fixed point } \label{sec:rev}

In this introductory section we recapitulate the physics of the EBL from a perspective that will be useful in subsequent sections. In all of what follows we will be working in a setting appropriate for doing condensed matter physics: we will be in (continuous) imaginary time, on a spatial lattice with a finite lattice spacing $a$, and will be performing all calculations in the thermodynamic limit. In this limit the system size $L$ is sent to infinity, while $a$ is held fixed. For more background information and a discussion of other types of starting assumptions, see e.g. \cite{paramekanti2002ring,tay2011possible,seiberg2003exotic,gorantla2021low}. Working in the thermodynamic limit is particularly important; if we were to instead use a continuum limit in which $a$ is sent to zero, our conclusions about stability would change, in line with the discussion of Ref. \cite{gorantla2021low}. We will have more detailed comments to make about this issue later on.

The EBL is a system of bosons on a 2+1D square lattice at average density $\ob n$. In most of what follows we will take $\ob n$ to be some generic (incommensurate) value. The dynamics of the bosons is assumed to be dominated by an onsite repulsion and a ring exchange hopping term, with the most important terms in the microscopic Hamiltonian schematically of the form 
\bea \label{uvham} H & \sim \mck \sum_i  b(\bfr_i)b^\da(\bfr_i+a\uvx)b(\bfr_i+a\uvx+a\uvy) b^\da(\bfr_i+a\uvy)+h.c. \\ & \qq + \mcu \sum_i (n(\bfr_i)-\ob n)^2  ,\eea 
where $n(\bfr_i)$ is the boson number operator, $\bfr_i$ runs over lattice sites, and $a$ is the lattice spacing. 
%We will be interested in the limit where $\mck/\mcu$ is large, so that the system is gapless and so that the ring-exchange term dictates how we approach dealing with the low energy physics. 

One important feature of the ring-exchange term is that it separately conserves the number of bosons along every row and column of the lattice. In the absence of any other boson hopping terms in the Hamiltonian such as\footnote{Here and in the following we will be abusing notation by letting $a$ stand for both the lattice spacing, and, when appearing in a sum, a coordinate index $a\in \{x,y\}$.} $\sum_a b^\da(\bfr_i) b(\bfr_i+a\uva)$, the Hamiltonian thus possesses a gigantic group of subsystem symmetries, with the boson number along each row and column of the lattice being separately conserved. In what follows we will never include this subsystem symmetry as part of our microscopic symmetry group, since a microscopic boson Hamiltonian with this symmetry group requires a large amount of fine-tuning. Rather, we will always imagine that the Hamiltonian above includes terms with small bare coefficients which break the subsystem symmetry. Part of the task at hand is to determine whether or not such terms are relevant (in the technical sense). 
The actual microscopic symmetry group we will work with in this note will at most consist of overall $U(1)$ boson number conservation, translation symmetry, and the discrete symmetries of the square lattice. 
In fact, for the purposes of the points we are trying to make, none of these symmetries are essential, and we will eventually relax them in subsequent sections. 

An analysis of the problem defined by the UV Hamiltonian \eqref{uvham} proceeds by using a hydrodynamic description in terms of two compact fields $\phi,\t$, which keep track of fluctuations in the phase and density of the UV bosons, respectively \cite{paramekanti2002ring}. This is done in a manner very similar to what we would do when writing down a hydrodynamic description of interacting bosons in 1+1D. The legitimacy of this approach can be justified a posteriori by computing correlation functions using the $\phi,\t$ description and noting their quasi 1+1D character, as well as by the fact that the various ordered phases of the theory can be accessed from the hydrodynamic description by turning on appropriate cosines of $\phi$ and $\t$ (to be discussed later). 

In more detail, the hydrodynamic description works by performing a polar decomposition on the microscopic boson annihilation operator by writing 
\be \label{uv_boson_rep} b = e^{i\phi} \sqrt{a^2\ob n + \frac1\twp\De_x\De_y\t},\ee 
where $\De_a$ denotes the dimensionless lattice gradient and $\phi$ is a compact field keeping track of the boson phase. The subsystem symmetries referred to above (which will always be broken microscopically) act as $\phi(\bfr) \mt \phi(\bfr) + f(x) + g(y)$ for arbitrary functions $f,g$. $\t$ on the other hand is a field which keeps track of the fluctuations in the boson density. It is defined on the sites of the dual lattice, so that $\De_x\De_y\t$ at a lattice site $\bfr$ can be written out as $(\De_x\De_y\t)(\bfr) = \t(\wt\bfr) - \t(\wt\bfr+a\uvx) + \t(\wt\bfr+a\uvx+a\uvy) - \t(\bfr+a\uvy)$, where $\wt\bfr = \bfr - (a/2,a/2)$. 

As written in \eqref{uv_boson_rep}, $\t$ must be constrained so that $a^2\ob n + \frac1\twp \De_x\De_y \t$ has integer eigenvalues; as usual this constraint will be enforced softly in the low-energy theory by letting $\t$ run over all real values, and adding cosines like $\cos(q[\twp  \ob n xy +  \t])$ to the low-energy action, with $q\in \zz$. The reason for parametrizing the fluctuations in the density as $\De_x \De_y\t$ is because $\t$ then determines the density of quadrupolar ring-exchange configurations of bosons \cite{paramekanti2002ring}, which given the form of the Hamiltonian are the most important density fluctuations to keep track of. 
Note that this whole procedure is exactly analogous to what we would do when studying interacting bosons in 1+1D, with the only difference being that in the latter case we would replace \eqref{uv_boson_rep} with $b= e^{i\phi} \sqrt{a \ob n + \frac1\twp \De_x\t}$. 

As mentioned above, we will mostly be interested in scenarios where the microscopic symmetries of interest are those of boson number conservation and translation, together with $D_4$ point group symmetry. These act on the hydrodynamic fields as \cite{paramekanti2002ring}
\bea \label{symm_action} U(1)\, :\, & \phi(\bfr) \mt \phi(\bfr) + \l ,\qq \t(\bfr)\mt \t(\bfr)\\ T_\bfmu\,:\,& \phi(\bfr)\mt \phi(\bfr+\bfmu), \\ & \t(\bfr) \mt \t(\bfr+\bfmu) + \twp \ob n(\mu_xy + \mu_y x + \mu_x\mu_y) \\ 
D_4 \, : \, & \phi(\bfr) \mt \phi(R_{\pi/4}^n R_x^m \bfr),\\ &   \t(\bfr) \mt (-1)^{n+m} \t(R_{\pi/4}^n R_x^m\bfr),\eea 
where $\l\in [0,2\pi)$, $T_\bfmu$ is the operator which implements translation through the vector $\bfmu = (\mu_x,\mu_y)$,\footnote{The action of $T_\bfmu$ on $\t$ can be understood by requiring that the boson density $n = \ob n + \frac{a^2}\twp \De_x\De_y \t$ transform as a density under continuous translations, with $n(\bfr) \mt n(\bfr+\bfmu)(1+\D\cdot \bfmu)$ to first order in $\mu$. In fact as written above this property only holds for transformations for which $\D_x\D_y \mu^y = \D_x\D_y\mu^x=0$ (these translations are more easily represented due to the form of the derivatives in $\De_x\De_y\t$), but restricting ourselves to these transformations will be enough for the present purposes. 
%	, since the theory we are interested is only invariant under the symmetries of the square lattice. 
Note also that the transformation of $\t$ is nonlinear, with the $\ob n\mu^x\mu^y$ piece ensuring that $T_{\bfmu}T_{\bfnu}\t = T_{\bfmu+\bfnu}\t$.} and we have written a general element in $D_4$ in terms of a $\pi/4$ rotation $R_{\pi/4}$ and a reflection about the $x$ axis $R_x$. 

In terms of the hydrodynamic variables $\phi,\t$, the appropriate action for the Hamiltonian \eqref{uvham} can be written as 
\bea \label{uvaction} S & = \int d\tau\sum_i\, \Big( i\p_\tau \phi \(a^2\ob n + \frac1\twp\De_x \De_y \t\) + \\ & \qq \frac{R^2\vs}\twp (1-\cos(\De_x\De_y\phi)) + \frac{\vs}{\fpi R^2} (\De_x \De_y\t)^2 \Big) + \cdots,\eea 
where the first term is the hydrodynamic representation of $b^\dagger \p_\tau b$, the sum over $i$ is over the implicit position argument $\bfr_i$ of the $\phi,\t$ fields, and where we have chosen to parametrize the couplings in the Hamiltonian in terms of an energy $\vs$ and a dimensionless constant $R^2$.\footnote{We will find the above parametrization of the couplings in terms of $R^2,\vs$ to be most convenient in what follows. In terms of the standard notation used e.g. in \cite{paramekanti2002ring}, we have $\vs \sim \twp \ob n \sqrt{\mcu\mck}, \, R^2 = \ob n \sqrt{\mck/\mcu}$.}
In \eqref{uvaction} the terms in $\cdots$ include other subdominant boson hopping terms such as $b^\da(\bfr_i + a\uva)b(\bfr_i)+h.c. \xra{\sim}\cos(\De_a\phi)$, as well as cosines like $\cos(q[\twp \ob n x y + \t]), \cos(q[\twp \ob n x + \De_y \t])$ etc. (with $q\in \zz$), which as mentioned above arise from softly constraining the dimensionless boson number density to be integer-valued. The terms written down explicitly in \eqref{uvaction} preserve an additional subsystem symmetry which acts on the $\t$ fields (corresponding to the conservation of vortex number on each row and column of the lattice), but this symmetry is broken completely by the aforementioned cosines involving $\t$. 

As in the hydrodynamic analysis of interacting bosons in 1+1D, we can analyze the low energy behavior of this system by first assuming that the cosines of $\t$ are small enough to allow $\t$ to be integrated out via Gaussian integration, producing an effective action in terms of $\phi$ alone. The legitimacy of this step can then be determined a posteriori by using an RG scheme to determine the relevance of the appropriate cosines. Doing this, we then obtain\footnote{Here we are intentionally omitting the total derivative term $\int d\tau \sum_i \ob na^2 \p_\tau \phi$, which will not be important to keep track of in what follows.}
\be \label{uvphiaction} S = \frac{R^2}\twp \int d\tau \sum_i \( \frac1{2\vs} (\p_\tau\phi)^2 + \vs (1-\cos(\De_x\De_y\phi))\)+\cdots,\ee 
where the $\cdots$ again contain all the nonlinear interactions allowed by symmetry and compactness of $\phi$. 

\begin{figure}
	\centering
	\includegraphics[width=.45\tw]{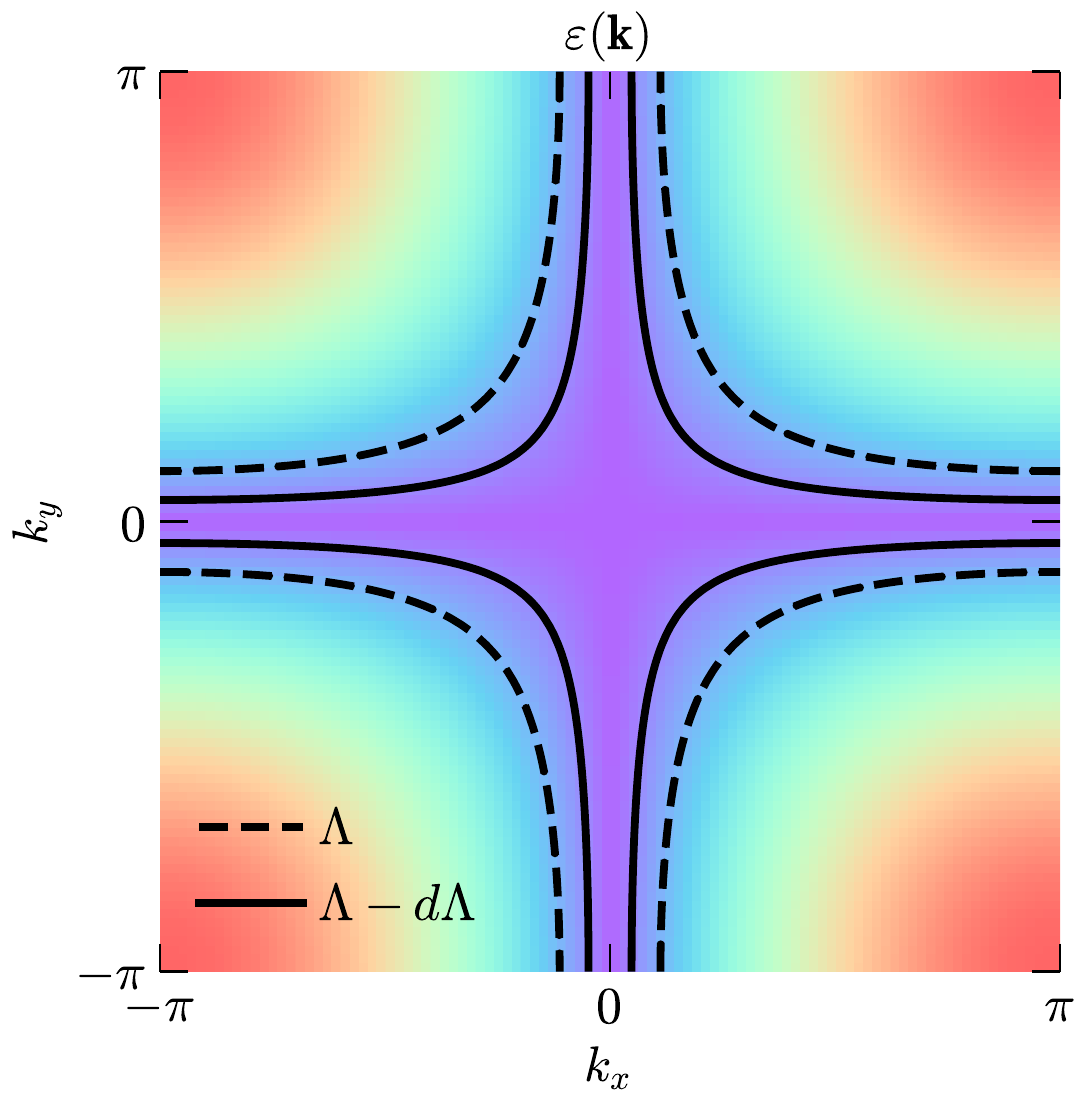}
	\caption{\label{fig:rg_fig} A color plot of the EBL dispersion, which is minimized along the coordinate axes. The dashed black contour indicates a cutoff at the momentum scale $\L\ll 1/a$; the solid black contour indicates one at $\L-d\L$. The modes lying between the two contours are integrated out during an RG step. }
\end{figure}

Since we are assuming that the dynamics of the bosons is dominated by the ring-exchange term, the modes which are relevant for describing the low energy physics are those for which $\De_x \De_y \phi$ (but {\it not} necessarily $\De_a\phi$) is small. This allows us to Taylor expand the cosine in \eqref{uvphiaction} to leading order, producing a quadratic action for $\phi$. Doing this, Fourier transforming to momentum space, and then generalizing by letting $R^2,\vs$ become nontrivial functions of momentum, the quadratic part of the action is then
\bea \label{general_gaussian_action} S_0 = \int_{\bfk,\o}   \frac{R^2(\bfk)}{\fpi\vs_\bfk} \( \o^2   + \ep_\bfk^2 \)|\phi(\o,\bfk)|^2, \eea 
 where $\int_{\bfk,\o} = \int \frac{d^2k}{(\twp/a)^2} \frac{d\o}{\twp}$, and where the dispersion is 
\be \label{disp} \ep_\bfk = 4 \vs_\bfk |\sin(k_xa/2) \sin(k_ya/2)|.\ee 
This dispersion is shown in fig. \ref{fig:rg_fig}, with its most salient feature being the fact that it vanishes along the coordinate axes in momentum space. 

Alternatively we may integrate out $\phi$ instead, producing a quadratic action for $\t$ that reads\footnote{The fact that $(\De_x\De_y)^2$ is not invertible doesn't really matter, as the zero modes can be fixed away by boundary conditions (just as in the case of inverting the Laplacian). }
\bea \label{theta_action} \int_{\bfk,\o} \frac1{\fpi R^2(\bfk)\vs_\bfk} (\o^2+\ep_\bfk^2) |\t(\o,\bfk)|^2.\eea 
The duality between $\phi$ and $\t$ 
%under which e.g. $\p_\tau \t / \vs  \lra \De_x \De_y \phi$ 
is much the same as in the 1+1D case, and simply sends $R^2(\bfk) \lra R^{-2}(\bfk)$. This free model, where all allowed cosines of $\t$ are neglected (corresponding to removing field configurations of $\phi$ containing vortices), is essentially the same as the modified Villain XY-plaquette model introduced in \cite{gorantla2021modified}.

The functional form of $\vs_\bfk$ will turn out to not affect any RG eigenvalues or other quantities of interest for the present discussion; therefore in what follows we will simply assume that $\vs_\bfk = \vs$ is independent of momentum. 
$R^2(\bfk)$ on the other hand does affect RG eigenvalues, and the physics described by \eqref{general_gaussian_action} depends in an essential way on its functional form. In what follows we will allow $R^2(\bfk)$ to be an arbitrary positive-definite function that is smooth on momentum scales of order $1/a$, so that the Fourier transform of $R^2(\bfk)$ is local in real space.

\section{RG procedure} \label{sec:rg}

In this section we set up an RG scheme that can be used to address the stability of the Gaussian action \eqref{general_gaussian_action}. The general approach we will take to RG in the EBL is inspired by Shankar's treatment of RG in Fermi liquids \cite{shankar1994renormalization}, and will essentially follow the procedure worked out in \cite{lake2021bose,lake2021fermi}.

%The data specifying the Gaussian fixed point are the action as above, together with a scheme for imposing a cutoff and performing RG. 
As mentioned in the introduction, the point of an RG analysis is to isolate the universal physics, regardless of the length scales involved. In the present setting, the universal physics of the fixed point is determined by the modes living near the coordinate axes in momentum space; small modifications to the dispersion in regions far away from the coordinate axes should therefore be counted as irrelevant within any useful RG scheme. We therefore impose a cutoff $\L$ in momentum space by restricting to modes of $\phi$ and $\t$ with momentum such that 
\be \label{cutoff}\ep_\bfk \leq \vs \eta^2, \ee 
where we have defined the small parameter 
\be \eta \equiv a\L\ll 1.\ee 
An illustration of the cutoff imposed by \eqref{cutoff} is shown in fig. \ref{fig:rg_fig}.
The fact that $\eta\ll 1$ means that $\De_x \De_y\phi$ is always small, in accordance with our assumption that we may get away with Taylor expanding the cosine in \eqref{uvphiaction}.

The perturbations to the fixed point we will need to consider in our stability analysis can all be expressed as\footnote{When $e^{i\mco}$ has long-range order, we will instead write the summand as $1-\cos(\mco(\bfr_i,\tau))$. } 
\be \label{deltas} \d S = g\eta^{d_\mco} \vs \int d\tau \, \sum_i \cos(\mco(\bfr_i,\tau)),\ee 
where $\mco$ is some equal-time polynomial in the $\phi,\t$ fields, $d_\mco$ is a number whose determination will be discussed shortly, and $g$ a small dimensionless coupling, of order $\eta^0$. Note that, as can be shown using the scheme developed in this section, other terms involving time derivatives such as $(\p_\tau \De_x \phi)^2$ are either marginal and go into modifying the function $R^2(\bfk)$, or else are irrelevant. Therefore in the following we will only consider terms without time derivatives. 

Even though the combination $g \vs \int d\tau$ is dimensionless, the cutoff explicitly makes an appearance in \eqref{deltas} by way of the factor $\eta^{d_\mco}$. 
%Such $\L$ dependence is made possible by the fact that the scale $a$ is part of the low-energy theory --- in a conventional scale-free fixed point the cutoff used as a regulator is the only length scale, and as such cannot appear in the way above. 
Determining the correct value to take for $d_{\mco}$ can be done by requiring that when evaluated on typical field configurations, the integrand in \eqref{deltas} is of the same order in $\eta$ as the kinetic term in the Gaussian fixed point action, viz. of order $(\De_x \De_y \phi)^2 \sim \eta^4$.\footnote{One can also determine $d_\mco$ by requiring that the perturbation $\d S$ make contributions to correlation functions / the free energy which are the same order in $\eta$ as the appropriate quantities in the free theory.} If $d_\mco$ is chosen to be larger than this value then $\d S$ is too small to have any affect as a perturbation, while if $d_\mco$ is chosen to be smaller than the effects of $\d S$ are not perturbatively small, which contradicts our assumptions. As we will see, $d_\mco$ essentially determines the effective dimension that the ``scaling'' dimension of $\mco$ is to be compared to when determining the relevance of $\d S$ (scare quotes added here as we will not really be doing any scaling per se---more on this shortly). Note that this effective dimension cannot simply be determined by the number of spacetime dimensions in which $\cos(\mco)$ has power-law correlations---as was suggested in \cite{paramekanti2002ring,tay2011possible}---since the nature of the problem means that our RG scheme cannot involve any uniform rescaling of spacetime (real-space conformal perturbation theory is rendered unworkably awkward for the same reason).  

The correct value to take for $d_\mco$ is determined on an operator-by-operator basis. As an example, for typical field configurations in the cutoff theory we have 
\be \label{ddphi} \De^n_{sx}\De^m_{py} \phi \sim \eta^{2\min(n,m)},\ee 
where $\De_{sa}\phi(\bfr) = \phi(\bfr) - \phi(\bfr + sa\uva)$, $s\in \zz$.
Thus in order to have $\eta^{d_{ \De^n_{sx}\De^m_{py} \phi}}(1-\cos( \De^n_{sx}\De^m_{py} \phi)) \sim (\De_x\De_y \phi)^2 \sim \eta^4$ on typical field configurations, we must take 
\be d_{\De^n_{sx}\De^m_{py}\phi} = 4(1-\min(n,m)).\ee 
Therefore e.g. $d_{\De_a\phi} = 4$, while an operator $\mco$ which is invariant under the row / column subsystem symmetries acting on $\phi$ (which can be written as \eqref{ddphi} with both $n,m$ nonzero) has $d_{\mco} \leq 0$.

Now we move on to the determination of RG eigenvalues. In each RG step, we first split up $\phi = \phi_< + \phi_>$ into fast and slow modes (and likewise for $\t$), with $\phi_>$ only containing modes satisfying
\be \label{momentumshell} (\eta')^2 < \ep_\bfk/\vs < \eta^2,\ee 
and with $\phi_<$ containing the rest, where we have defined
\be \eta' \equiv \eta (1-dt)\ee 
with $dt = d\L / \L \ll1$ the RG time step.
Note that the modes being integrated out include modes of all frequencies --- given the non-relativistic nature of the problem and the lack of a need for a frequency cutoff when calculating correlation functions, it is more natural to simply integrate out all frequencies, thereby keeping the effective action local in time. 

Integrating out the fast fields, to lowest order in $g$ the perturbation to the slow field effective action is  
\be \label{dsint}\d S = g \vs \eta^{d_\mco} \int d\tau \sum_i \cos(\mco_<) e^{-\frac12 G_{\mco_>}(0,\bfzero)}\ee 
where $G_{\mco_>}$ is the 2-point correlation function of $\mco_>$, with the decomposition $\mco=\mco_<+\mco_>$ induced from those of $\phi$ and $\t$. We then define the ``scaling'' dimension $\De_\mco$ by 
\be \label{sddef} G_{\mco_>}(0,\bfzero) = 4\De_\mco dt + O(dt^2).\ee
The factor of $4$ here (a factor of 2 would be more normal) arises because with this definition the power laws that appear in the correlation functions of $\mco$ are functions of spacetime distances to the power of $2\De_\mco$ (see appendix \ref{sec:app}).
Rewriting the $\eta^{d_\mco}$ appearing in \eqref{dsint} in terms of $\eta'$, we then see that the mode integration effectively results in $g$ being replaced by 
\be g' = g(1+(d_\mco-2\De_\mco)dt),\ee so that the RG eigenvalue of $g$ is 
\be y_g = d_\mco-2\De_\mco.\ee 
Therefore whether or not $g$ represents a relevant perturbation is determined by comparing $2\De_\mco$ to the effective dimension $d_\mco$.

Note that at no point have we rescaled coordinates so that the cutoff is increased back to $\L$; we have instead simply re-expressed $\eta$ in terms of the new (reduced) cutoff. Due to the form of the dispersion a re-scaling which returns the cutoff to its original value cannot be uniform in momentum space, and as such must necessarily have a rather nasty implementation in real space, which is where the hydrodynamic description of the EBL is most naturally formulated (moreover, any such rescaling is ultimately nothing more than a change of variables, and cannot by itself contain any physical content).

\section{Stability analysis} \label{sec:stab}

We now apply the general discussion of the preceding section to compute scaling dimensions of operators in the EBL theory, with the goal of determining the stability of the free fixed point \eqref{general_gaussian_action}.

\ss{``Scaling'' dimensions of operators} 

Before committing to a particular choice for the function $R^2(\bfk)$, let us make a few general comments. 
The operators added as perturbations to the free fixed point that we are interested in can all be written as either $\cos(\cp_q)$ or $\cos(\ct_q)$, where $\cp_q$, $\ct_q$ represent general integral linear combinations of $\phi,\t$ fields, respectively:
\be \label{cpctdef}\cp_q = \sum_i q_i \phi(\bfr_i),\qq \ct_q = \sum_i q_i \t(\bfr_i),\qq q_i\in \zz.\ee 
The scaling dimensions of $\cos(\cp_q),\ \cos(\ct_q)$ can be computed as follows. 
First, consider $\cos(\cp_q)$ operators. 
Letting $dS_\L$ denote the shell in momentum space containing fast modes with momenta satisfying \eqref{momentumshell}, the scaling dimension of $\cos(\cp_q)$ is extracted by computing the fast-mode correlator as (working to leading order in $dt$)
\bea \label{scaling_integral} 4dt \De_{\cp_q} & =  \int_{d S_\L} \frac{dk_x dk_y}{(\twp/a)^2} \int_\rr d\o \,  \frac{|q(\bfk)|^2}{R^2(\bfk)} \\ & \qq \times \frac1{\o^2/\vs + \vs (4|\sin(k_xa/2) \sin(k_ya/2)|)^2} \\ 
& =  \frac1{\fpi \L^2} \int_{dS_\L} dk_x \, dk_y\,  \frac{|q(\bfk)|^2}{R^2(\bfk)},\eea 
with $q(\bfk)$ the lattice Fourier transform of $q_i$. 

 As mentioned above, $R^2(\bfk)$ is assumed by spatial locality to be the Fourier transform of a function which is localized on length scales below $1/\L^2 a = 1/\eta \L$. 
This means that $R^2(\bfk)$ can be Taylor expanded about zero momentum when either $k_x$ or $k_y$ is much less than $\pi/a$, in particular when $k_x,k_y \sim \L$. 
We furthermore will only be interested in operators $\cp_q$ which are themselves local on scales below $1/\eta \L$, so that $|q(\bfk)|^2$ can be similarly expanded. 
Now the integral over the shell $dS_\L$ can be split into regions where $k_xa\ll1$ and $k_y$ ranges from $\L$ to $\pi/a$ (for which the shell $d S_\L$ is defined via $k_x = \L^2 a/2\sin(k_y a/2)$), and likewise for $k_x\lra k_y$. Performing appropriate Taylor expansions of $R^2(\bfk)$ and $q(\bfk)$ in these regions, 
we may perform the integral over $dS_\L$ and write 
\bea \label{dcoscp}\De_{\cp_q}
& = \frac{a}{4\pi} \int_\L^{\pi/a}\Big( \frac{dk_x\, |q(k_x,\L^2 a/2\sin(k_xa/2))|^2}{R^2(k_x,\L^2 a/2\sin(k_xa/2)) \sin(ak_x/2)} \\ & \qq  +\frac{dk_y\, |q(\L^2 a/2\sin(k_ya/2),k_y)|^2}{R^2(\L^2 a/2\sin(k_ya/2),k_y) \sin(ak_y/2)}\Big) \\ 
&\approx  \frac a{4\pi}  \int_\L^{\pi/a}\Big( \frac{dk_x\, |q(k_x,0)|^2}{R^2(k_x,0) \sin(ak_x/2)} \\ & \qq +\frac{dk_y\, |q(0,k_y)|^2}{R^2(0,k_y) \sin(ak_y/2)}\Big), \eea
where $\approx$ means equality up to terms suppressed by higher powers of $\eta$.
  
There are several things to note about this expression. First, note that the scaling dimension depends only on the values that $R^2(\bfk)$ takes on the coordinate axes. Therefore for the purposes of determining the stability of the Gaussian fixed point we only need to know the function $R^2(k_x,0)$ (which is equal to $R^2(0,k_y)$ by the square lattice symmetry we have assumed to be present for simplicity). 

Secondly, note that $\De_{\cp_q}$ diverges logarithmically as $-\ln(\eta)$ unless $q(\bfzero)=0$, since $R^2(\bfzero)$ is a assumed to be finite. This implies that $\De_{\cp_q}$ diverges if $\sum_i q_i \neq 0$, i.e. if $\cp_q$ is charged under the global $U(1)$ boson number symmetry. Hence only operators which conserve total boson number have a chance to be relevant.
%This divergence is just a manifestation of the fact that such operators have ultra short ranged correlations at the EBL fixed point \cite{paramekanti2002ring} (although note that it is a UV divergence, i.e. one which does not rely on taking the thermodynamic limit). 

Thirdly, note that the scaling dimension vanishes with $\eta\ra0$ if $q(k_x,0) = q(0,k_y) = 0$ for all $k_x,k_y$. This condition is equivalent to the condition that $\cp_q$ be neutral under the row and column subsystem symmetries which act on $\phi$. As was discussed near \eqref{ddphi}, any operator neutral under both row and column subsystem symmetries must have $d_\mco \leq 0$. Together with the fact that such operators have $\De_\mco = O(\eta)\ra 0$, this means that any perturbation preserving the subsystem symmetries is guaranteed to be either irrelevant or marginal. 

 Therefore any operator $\cos(\cp_q)$ which has a chance to be relevant must both a) preserve the global $U(1)$ symmetry and b) break the subsystem symmetries. From the discussion near \eqref{ddphi} we see that such operators have $d_\mco = 4$; as such their relevance is determined by comparing their scaling dimensions to 2.
% , and can be ignored for the purposes of this section. 
 
 Above we have focused on $\cos(\cp_q)$ operators. The story for the $\cos(\ct_q)$ operators is the same, with the only difference in the calculation of $\De_{\ct_q}$ being the replacement $R^2(\bfk) \ra R^{-2}(\bfk)$. In particular, any operator with nonzero vortex number (such as $\cos(\t)$) will be infinitely irrelevant, while any operator invariant under the dual subsystem symmetries (which count the vortex number in each row and column of the lattice) will be either irrelevant or marginal.

From \eqref{dcoscp} is clear that the operators with the smallest nonzero scaling dimensions must either have $q(k_x,0) = 0$ for all $k_x$ or $q(0,k_y) = 0$ for all $k_y$, but not both. Therefore when going about finding operators which have the potential to destabilize the fixed point, we may restrict our attention to operators which break one of the row / column subsystem symmetries, but not both.
%\footnote{To see this, consider a general neutral configuration of charges. Consider re-arranging the charges so that they all have the same $y$ coordinate, without changing their $x$ coordinates. After this re-arrangement, the new charge configuration has $q(0,k_y) = 0$ for all $k_y$, while $q(k_x,0)$ is unchanged. Therefore this re-arrangement either decreases the scaling dimension or leaves it unchanged. As such, the operators with the minimal scaling dimensions can always be taken to have either $q(0,k_y)=0$ for all $k_y$, or $q(k_x,0) = 0$ for all $k_x$.} 
Without loss of generality we may thus only consider operators with nonzero $q(k_x) \equiv q(k_x,0)$. Combined with the fact that the scaling dimensions of such operators depend only on $R^2(k_x,0)$, the calculation of the smallest scaling dimensions appearing in the operator spectrum reduces to a one-dimensional optimization problem, considerably simplifying the stability analysis.

%The question of stability can therefore be framed as follows. Consider the quadratic forms\footnote{Here we can taken the lower limit of the integral to be 0 instead of $\L$ as in \eqref{dcoscp}; this can be done as long as $q(\bfzero)=0$, which will be true for all the operators we are interested in.}
%\bea \label{general_quadratic_forms} Q_\cp(q,q') & = \frac1{\fpi} \int_0^{\pi/a} \frac{dk}{R^2(k_x,0) \sin(ka/2)}  q(k)^* q'(k) \\ 
%Q_\ct(q,q') & = \frac1{\fpi} \int_0^{\pi/a} \frac{dk}{\sin(k_xa/2)} R^2(k_x,0) q(k)^*q'(k), \eea 
%where $q(k) = \sum_j e^{-ikj}q_j$, $q_j\in \zz$, with $j$ running over sites of a one-dimensional lattice (and likewise for $q'(k)$), and where $\{q_j\}$ . 
%These quadratic forms define two lattices $\mcl_\cp = \sqrt{Q_\cp} \zz^L$, $\mcl_{\ct} = \sqrt{Q_\ct} \zz^L$, with $L\ra\infty$ the system size. The Guassian fixed point parameterized by the given choice of $R^2(\bfk)$ is then stable provided that neither of these two lattices contain a point within a distance of 2 of the origin. 
%
When performing a numerical search for relevant operators, it is helpful to further simplify things slightly. Since all $\cp_q$ ($\ct_q$) operators we need to consider have zero (vortex) charge, implying that $q(0) = 0$, we may without loss of generality write $q(k)$ in terms of integers $q_l$ defined on the links of a one-dimensional lattice as
\be \label{qk} q(k) = \sum_{l\in\zz} e^{-ikla} q_l (1 - e^{-ika}).\ee
In terms of the $q_l$, the scaling dimensions are 
\bea \label{simple_sds} \De_{\cp_q} & = \frac1\pi \int_0^{\pi/a} dk\, \frac{\sin(ka/2)  }{R^2(k_x,0)}\sum_{l,l'}q_lq_{l'} \cos([l-l']ka) \\ 
\De_{\ct_q} & = \frac1\pi \int_0^{\pi/a} dk\, R^2(k_x,0) \sin(ka/2) \sum_{l,l'}q_lq_{l'} \cos([l-l']ka).\eea 
The Guassian fixed point parameterized by the given choice of $R^2(\bfk)$ is then stable provided that there are no nontrivial choices of $\{q_l\}$ for which at least one of $\De_{\cp_q},\De_{\ct_q}$ is less than $2$.
 
For now, we will only be interested in perturbations which respect translation symmetry. From the translation action of \eqref{symm_action}, it is easy to check that at generic incommensurate $\ob n$,
%a linear combination $\ct_q$ of $\t$ fields transform as 
%\be T_\bfmu \,:\, \ct_q(\bfr) \mt \ct_q(\bfr+\bfmu) + \bfD(q)_x \mu_y + \bfD(q)_y \mu_x + M(q)\mu_x\mu_y,\ee 
%where $\bfD(q)$ is the (vortex) dipole moment of $\ct_q$, viz. $\bfD(q) = \sum_a q_i \bfr_i$. 
any translation-invariant operator $\cos(\ct_q)$ must have zero vortex dipole moment\footnote{This statement applies to cosines of integral linear combinations of $\t$ fields. One may also use explicit coordinate dependence to produce translation invariant cosines, e.g. as $\cos(\twp \ob n ax - \De_y \t)$. However as any finite-action field configuration must asymptotically have $\De_x\De_y\t\ra 0$, such cosines are always rapidly oscillating at large distances for incommensurate $\ob n$, and as such can be ignored \cite{paramekanti2002ring}.} (such as e.g. $\cos(\De_{a}^2\t)$). 
 Therefore translation invariance allows us to restrict the $\{ q_l\}$ in the second line of \eqref{simple_sds} to those integers satisfying $\sum_l q_l = 0$.

\ss{Constant $R^2(\bfk)$}

Let us first consider the case where $R^2(k_x,0) = R^2$ is a constant, independent of momentum. In this case there turns out to always exist a symmetry-allowed relevant perturbation. 

To show this, we start by considering the simplest $\cp_q$ operators which preserve the total boson number, viz. exponentials of $\De_x\phi$ and $\De_y\phi$. The dimension of these operators is, reading off from \eqref{simple_sds}, 
\be \De_{\De_x\phi} = \frac1{\pi R^2} \int_0^\pi dx\, \sin(x/2) = \frac{2}{\pi R^2}.\ee 
As $d_{\De_a\phi} = 4$ (so that $\De_{\De_x\phi}$ should be compared to 2 when determining relevance), stability of the EBL fixed point with constant $R^2$ requires 
\be\label{phi_irrel} R^2 < 1/\pi.\ee 

On the other hand, consider translation-invariant operators $\ct_q$ built from the $\t$ fields. The simplest such operators are those involving two derivatives, which may be written as $\De_{sx}\De_{px} \t$ for some integers $s,p\neq 0$.
%\footnote{The definition of $\De_{sx}$ is below \eqref{ddphi}; concretely, $(\De_{sx}\De_{px} \t)(r) = \t(r) - \t(r+p) - \t(r+s) + \t(r+p+s)$.}
These operators have scaling dimensions 
\be \De_{\De_{sx}\De_{px} \t} = \frac{4R^2}\pi \int_0^\pi dx\, \frac{\sin^2(sx/2) \sin^2(px/2)}{\sin(x/2)}.\ee 
The RHS is minimized when one of $s,p$ is equal to unity, with the other made large. Specifically, if we set $p=1$ (such operators were previously identified in \cite{tay2011possible}), we find 
\be \De_{\De_{sx} \De_x \t} = \frac{4R^2}{\pi(1-1/4s^2)}.\ee 
Strictly speaking, the most relevant operator is therefore $\De_{(s\ra\infty)x} \De_x \t$. However, on general grounds the bare coupling constant for $\cos(\De_{(s\ra\infty)x} \De_x \t)$ will decay (usually exponentially fast) with $s$. Since the variation of the scaling dimension with $s$ is rather small, we will simply restrict our attention to $s=1$, with the scaling dimension 
\be \De_{\De_x^2\t} = \frac{16R^2}{3\pi}.\ee 
This operator is therefore only irrelevant provided that 
\be \label{theta_irrel}R^2 > \frac{3\pi}{8} .\ee 

By examining \eqref{phi_irrel} and \eqref{theta_irrel}, we see that no matter the value of $R^2$, there is always a relevant perturbation. We now briefly discuss the nature of the RG flows driven by these perturbations. 

\sss{$\cos(\De_a\phi)$ most relevant}

If $\cos(\De_a\phi)$ is the most relevant perturbation and a term like $\sum_a[1-\cos(\De_a\phi)]$ is added to the action, the flow drives the model to a regime where $\De_a\phi\ll1$ on typical field configurations, so that the aforementioned cosine can be Taylor expanded. The resulting $\sum_a(\De_a\phi)^2$ term lifts the degeneracy of the dispersion, and in the IR we obtain the ordered phase of the 2+1D XY model. In particular, the IR theory is massless, and the response to a background $U(1)$ gauge field is that of a superconductor (unlike the response of the EBL phase described by \eqref{general_gaussian_action}, which is insulating \cite{paramekanti2002ring}). 

\sss{$\cos(\De^2_a\t)$ most relevant}

Now consider the case where $\cos(\De^2_a\t)$ is the most relevant perturbation. As was discussed above, technically speaking $\cos(\De_{sa}\De_a \t)$ is more relevant for larger $s$, but will also appear with a bare coupling constant that is suppressed rather quickly with large $s$. Since the effects of these operators for different small values of $s$ are similar, we will simply assume that the most important operator for determining the RG flow is $\cos(\De_a^2\t)$. These operators break down the group of subsystem symmetries acting on $\t$ in \eqref{theta_action} to the subgroup corresponding to conservation of total vortex charge and total vortex dipole moment (aka momentum). 

In this case, when a term like $\sum_a [1-\cos(\De^2_a \t)]$ is added to the action, we flow to a regime where $\De_a \De_b \t \ll 1$ for all $a,b = x,y$. After Taylor expanding, the free part of the action can then be written in continuum notation as  
\be \label{qlm} S_0 = \int d\tau \, d^2x\, \( (\p_\tau \t)^2  + A\sum_a (\p_a^2 \t)^2 + B (\p_x\p_y\t)^2\) \ee 
for constants $A,B$. This is essentially the quantum Lifshitz model with a square lattice anisotropy,\footnote{The usual quantum Lifshitz model on the square lattice is generically unstable \cite{vishwanath2004quantum,fradkin2004bipartite} due to nonlinear terms involving derivatives like $(\partial_x\theta)^4 + (\partial_y\t)^4$. In the present case such terms are disallowed by translation invariance. } with the added proviso that, by translation invariance, all terms must preserve the vortex dipole moment (viz. must either involve time derivatives, or at least two spatial derivatives). There is no longer any degeneracy of the dispersion along the coordinate axes. However, since the terms in \eqref{qlm} are all quartic in spatial derivatives, $e^{i\t}$ has spatial correlation functions going schematically like $\sim \exp\( - \int d^2k \,  \frac{1-\cos(\bfk\cdot\bfr)}{k^2}\)$, and therefore cannot develop long-range order. On the other hand, the operator $e^{i\De_a\t}$ {\it does} have long-range correlators, and consequently develops long-range order, spontaneously breaking the conservation of vortex dipole moment. Since $\De_a \t$ is charged under translation, the resulting state spontaneously breaks translation symmetry, and possesses a gapless phonon mode. That this phase is gapless is in keeping with a general analysis of the t' Hooft anomalies present in the fixed point action \eqref{general_gaussian_action}, with an anomaly being present unless either one subsystem symmetry is fully broken, or both subsystem symmetries are broken down to their global $U(1)$ subgroups \cite{hotat}.

% a 
The exact type of charge ordering that occurs can be determined by writing down a low-energy expression for the density operator. In the microscopic boson model, the fluctuations in the number density are given by $\d n = \frac1{\twp a^2} \De_x \De_y \t$, as written down in \eqref{uv_boson_rep}. By the time enough high-energy degrees of freedom are integrated so as to land us in the low-energy hydrodynamic description, this expression will be generically modified to include all other terms involving $\t$ which are allowed by symmetry. The constraints of $D_4$ symmetry and the particle-hole symmetry occurring at half-filling mean that we may write (see e.g. \cite{tay2011possible,xu2007bond} for related discussions)
\bea \d n(\bfr) & = \frac1{\twp a^2} (\De_x \De_y \t)(\bfr) \\ &  + \sum_{q\in \nn} \Upsilon_q \De_x \De_y \sin(q[\t + \twp \ob n x y ]))(\bfr) \\ &+ \sum_{p\in \nn}  \G_p \((\De_x \sin(p[\De_y \t + \twp \ob n a x]))(\bfr) + (x\lra y)\)\\ & +  \cdots,\eea 
where the $\Upsilon_q,\G_p$ are non-universal coefficients, and the $\cdots$ represent subleading contributions to $\d n(\bfr)$ involving $\t$ fields living beyond the four dual lattice sites nearest to $\bfr$. 
In the ordered phase we are interested in, we have $\lan e^{i\De_a \t} \ran \neq 0$ and $\lan e^{i\t} \ran = 0$. In this case it is the components of the density involving $\G_p$ which order, and we may write  
\bea \lan \d n \ran & \approx  \twp \ob n a^2 \sum_{p\in \nn} p \G_p\big(\lan \cos(p\De_x \t)\ran \cos(\twp p \ob n ay)  \\ & \qq -  \lan \sin (p\De_x \t) \ran \sin(\twp p \ob n ay)\big) + (x\lra y). \eea

%We therefore conclude that the EBL with constant $R^2$ is always unstable with respect either to particle or vortex condensation, a conclusion which was also reached in \cite{tay2011possible} (although there without the support of a concrete RG scheme). 

\ss{General $R^2(\bfk)$}

We now turn to asking whether or not there exist more complicated choices of $R^2(k_x,0)$ that yield phases stable against the condensation of either particle or vortex dipoles. 

In general, this problem can be formulated as a shortest vector problem, where the task is to identify the length of the shortest vector in the integral lattices definable from \eqref{simple_sds}. This problem is generically very difficult (especially as the dimension of the lattice increases; here the dimension goes to infinity in the thermodynamic limit), and in the absence of more sophisticated constructive approaches like those of \cite{plamadeala2014perfect}, we must resort to a brute-force numerical search. This is ameliorated somewhat by the fact that a large number of choices for $\{q_l\}$ can be ruled out from the beginning (for example, we may require $\gcd(\{q_l\})=1$ and $\sum_l q_l\geq 0$ without loss of generality), but it still means that we will only at best be able to provide suggestive evidence for (but not prove) the existence of a stable phase. 

In the following, we perform our numerical search by restricting ourselves to operators with involve no more than $\sf{MaxBody}$ boson / vortex creation and annihilation operators, with the given operators separated by no more than $\sf{MaxRange}$ lattice points. To rigorously demonstrate stability, we would need to make arguments for what happens as $\mb, \mr \ra \infty$. Instead, we will simply content ourselves with performing a stability analysis for a series of increasing values of $\mb,\mr$, and observing whether or not any putative choices of $R^2(\bfk)$ appear to be stable as $\mb,\mr$ are increased. Note that if the system develops an instability only when $\mb$ or $\mr$ are increased past some large value, the EBL with the given choice of $R^2(\bfk)$ may be able to be regarded as stable in practice, as very high body operators / those that act over a large number of sites will generically enter the UV action with very small coefficients, preventing them from appearing until the very latest stages of the RG flow (before which in any real system the flow will be cut off either by finite size effects or by nonzero temperature). 

There are of course an infinite number of functions $R^2(k_x,0)$ to try when searching for a stable phase. One simple choice which seems to do a good job is
\be \label{rchoice} R(k_x,0) = \l_1(1+\l_2 \cos(ak_x))^2,\ee
with $\l_1>0$ and $|\l_2 |<1$. Note that this meets the requirements of being positive and being the Fourier transform of a reasonably localized function.

\begin{figure*}
	\centering 
	\includegraphics[width=.85\tw]{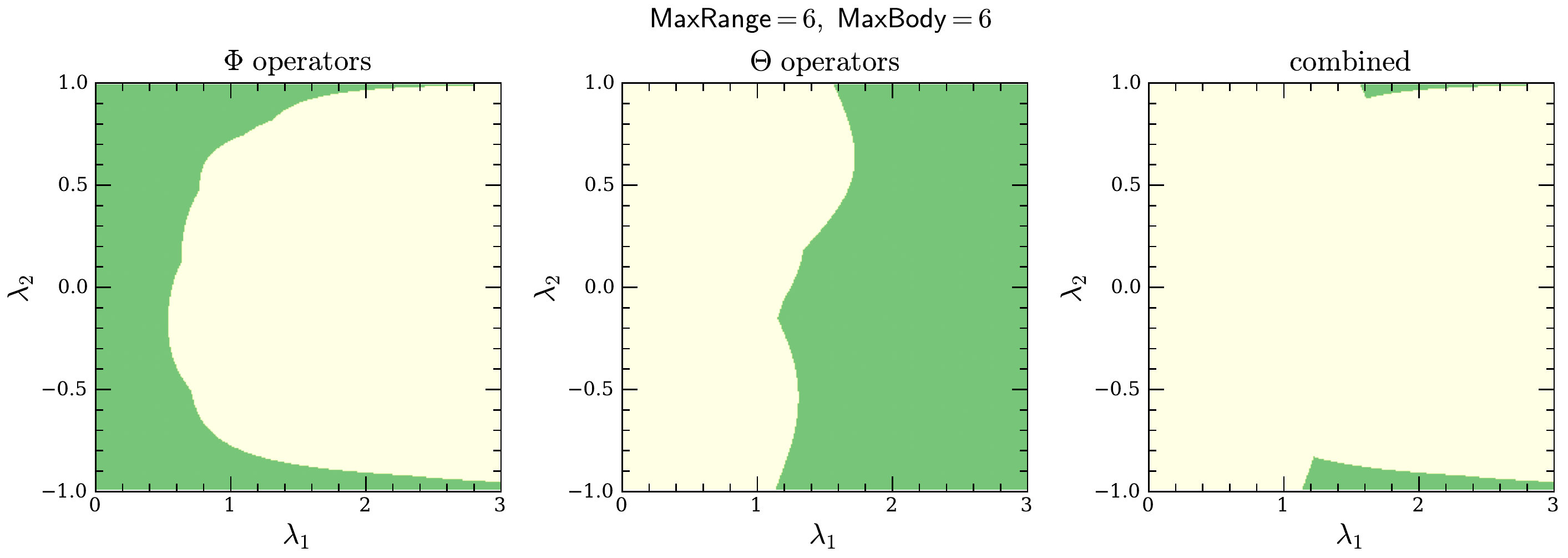}
	\caption{\label{fig:3panel_stab_fig} Stability of the EBL for $R^2(k_x,0)$ given by \eqref{rchoice} and $\mb=6,\mr=6$. Green regions indicate stability, i.e. the absence of relevant operators. Left: regions where no relevant $\cos(\cp_q)$ operators exist. Center: regions where no relevant $\cos(\ct_q)$ operators exist. Right: regions where no relevant operator of either type exists.}
\end{figure*}

A plot showing the regions in $\l_1,\l_2$ parameter space where \eqref{rchoice} yields a stable phase for $\mb , \mr = 6$ is shown in \ref{fig:3panel_stab_fig}. We see that there exist small regions of stability (green regions in the rightmost panel of fig. \ref{fig:3panel_stab_fig}), which are located near the regions where $|\l_2|=1$. The fact that the regions of stability are located near the border of the allowed parameter space is a common theme in these types of problems \cite{lake2021subdimensional,mukhopadhyay2001sliding,vishwanath2001two}. 

\begin{figure*}
		\centering 
	\includegraphics[width=.7\tw]{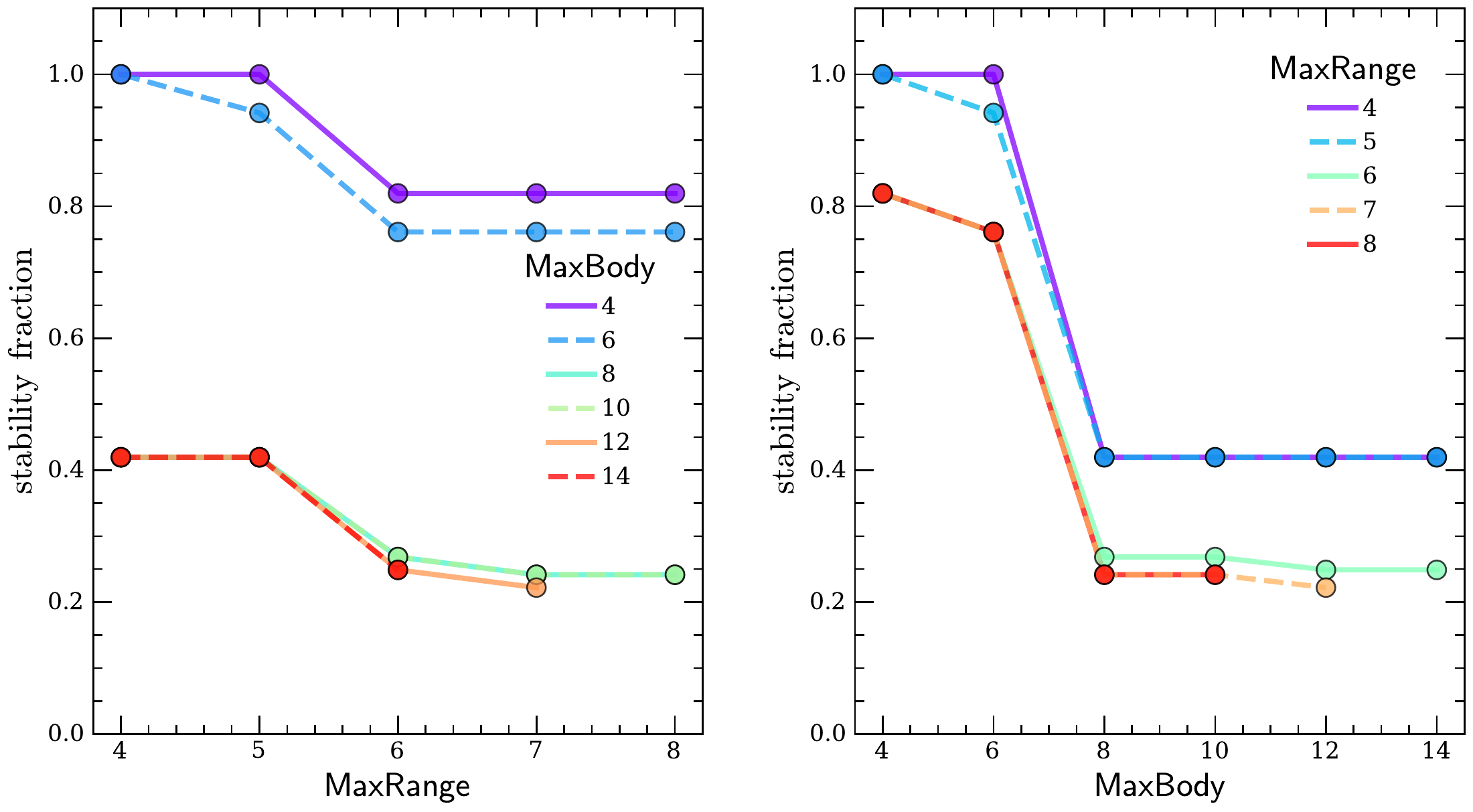}
	\caption{\label{fig:mbmr_fig} Size of the region of stability in Fig. \ref{fig:3panel_stab_fig} near $-1<\l_2<-0.75$, $1<\l_1<4$, as a function of $\mb$ and $\mr$. Here ``stability fraction'' denotes the area in $\l_1$-$\l_2$ parameter space of the stable region, relative to the area when $\mb=\mr=4$.}
\end{figure*}

Upon increasing $\mb,\mr$, the region of stability shrinks somewhat (especially at $\mb=8$), but does not completely disappear up to the largest values of $\mb,\mr$ we have numerically checked. The evolution of the size of the stable region with $\mb,\mr$ is shown in fig. \ref{fig:mbmr_fig}. 
Whether or not these curves should be viewed as extrapolating to a nonzero value in the $\mb,\mr \ra \infty$ limit is a decision left to the reader, but we remark again that even if the allowed region of stability vanishes when $\mb$ or $\mr$ are very large, the EBL in this parameter range may still be stable for all practical purposes. 

In the regions of instability, the exact RG flow will depend on the form of the most relevant operators, as well as the strengths of their bare coupling constants. In general though, we expect that the flow will be as in the constant $R^2$ case, viz. either towards the superfluid phase of the 2+1D XY model, or towards a translation-breaking crystalline state. 

\ss{Stability for other symmetry groups} 

So far we have restricted ourselves to perturbations which respect the symmetries of translation and total boson number conservation (as well as square lattice symmetry, though the latter is inessential). Boson number conservation is inconsequential to our stability analysis, since as we have seen operators which do not conserve boson number have infinite scaling dimensions. However, the stability analysis will indeed change if we relax our imposition of translation symmetry, or if we impose additional symmetries. 

\sss{No translation symmetry / commensurate density}

\begin{figure*}
	\centering
	\includegraphics[width=.85\tw]{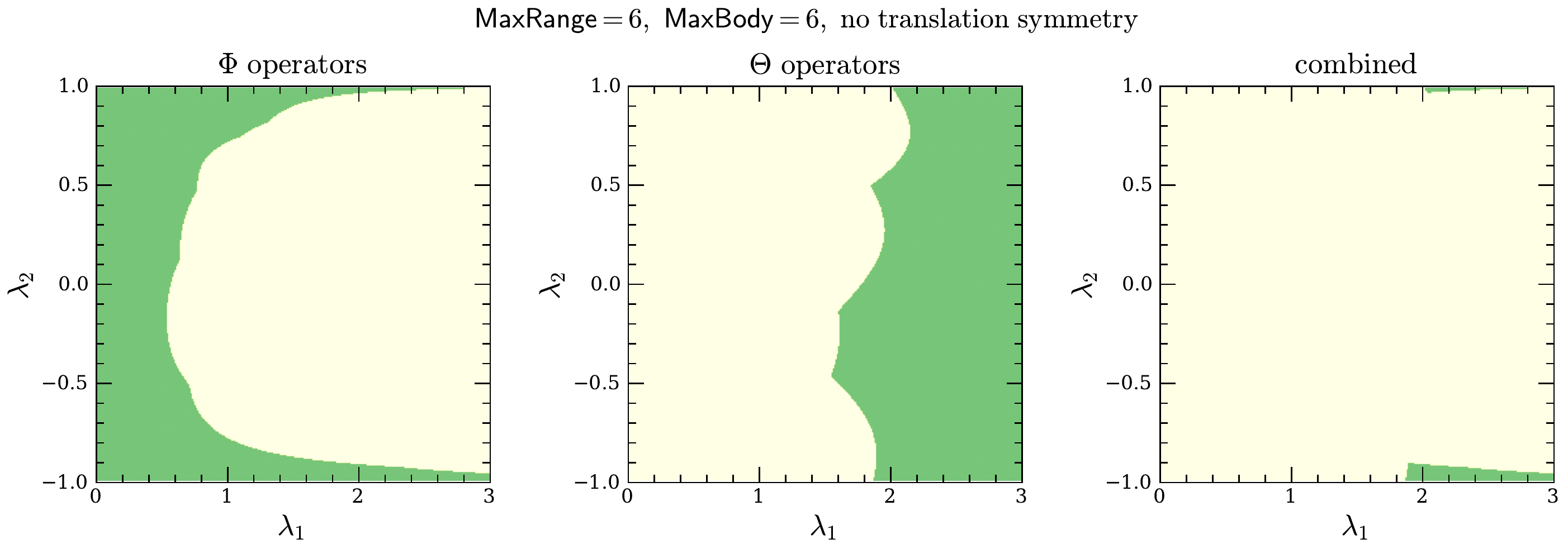}
	\caption{\label{fig:3panel_notrans_stab_fig} The same as in fig. \ref{fig:3panel_stab_fig}, but working at integral densities / including operators which break translation symmetry.}
\end{figure*}

Suppose now that we ignore the requirement that perturbations to the fixed point theory preserve translation symmetry. This then forces us to consider perturbations involving any combinations of $\t$ fields with zero vortex number\footnote{Operators with nonzero vortex number are still irrelevant, by the reasoning given earlier.} regardless of their dipole moment, such as e.g. $\sum_a\cos(\De_a \t)$. The same type of perturbations are allowed if we keep translation symmetry but work at a commensurate density with $a^2\ob n \in \zz^{\geq0}$, so that on average there are an integer number of bosons per site. In this case the factors involving explicit coordinate dependence which appear in cosines involving $\t$ fields (such as the $\twp \ob n x$ in $\cos(\De_y \t + \twp \ob n ax)$) can all be dropped on account of $\twp \ob n a x \in \twp \zz$ for all lattice sites $x$. From the perspective of stability, this case is therefore equivalent to the one where we ignore translation symmetry. 

In this setting, we may now consider arbitrary sets of integers $\{q_l\}$ in the second line of \eqref{simple_sds}. Obviously any region of stability in this case must be a proper subset of the region of stability found in the case where translation symmetry was imposed microscopically. We find that for the choice of $R^2(k_x,0)$ in \eqref{rchoice}, the region of (apparent) stability is reduced but not altogether eliminated, as shown in fig. \ref{fig:3panel_notrans_stab_fig}.

Consider the regions of instability where the RG flow is not towards the superfluid phase. If $a^2\ob n$ is not an integer, the flow will be towards some state with a pattern of charge order determined by the most relevant perturbation. If instead $a^2\ob n$ is integral, the flow will be towards a translation-invariant Mott insulator with $a^2 \ob n$ bosons per site. Either way, the resulting phase will be massive, which is allowed by anomaly constraints due to the dual vortex subsystem symmetry being completely broken \cite{hotat}.

 If translation symmetry is imposed and $a^2\ob n$ is not an integer but some relatively commensurate rational number, more cosines involving $\t$ operators are allowed, even in the presence of translation symmetry. The region of stability in this case will then be somewhere between the regions shown in figs. \ref{fig:3panel_stab_fig} and \ref{fig:3panel_notrans_stab_fig}, depending on the value of $a^2 \ob n$.  
In the regions of instability, if the most relevant operator is a cosine of $\t$, the resulting RG flow will generally be towards some sort of charge density wave; see \cite{paramekanti2002ring,tay2011possible} for a detailed discussion.

\sss{Dipole conservation} 

We may also consider a theory with a larger (but still finite-dimensional) global symmetry group. One symmetry we may impose is global dipole conservation, which maps $\phi \mt \phi + \a x +\b y$ for constant $\a,\b$, and under which operators like $\cos(\De_x\phi)$ carry charge. If we impose this symmetry in addition to translation, the region of stability in fig. \ref{fig:3panel_stab_fig} can only increase. 

Dipole conservation together with translation symmetry is however not enough to render the theory with constant $R^2$ stable. Indeed, irrelevance of $\cos(\De_x^2\t)$ still requires that $R^2>3\pi/8$, while the simplest dipole-neutral operator $\cos(\De_x^2\phi)$ has scaling dimension 
\bea \De_{\De_x^2\phi} & = \frac4{\pi R^2} \int_0^\pi dx\, \sin^3(x/2) = \frac{16}{3\pi R^2},\eea 
and as such is only irrelevant provided $R^2<8/3\pi$. Since $3\pi/8>1$, there is still no choice of constant $R^2$ for which both $\cos(\De_x^2\phi),\, \cos(\De_x^2\t)$ are irrelevant.  

\begin{figure*}
	\centering
	\includegraphics[width=.85\tw]{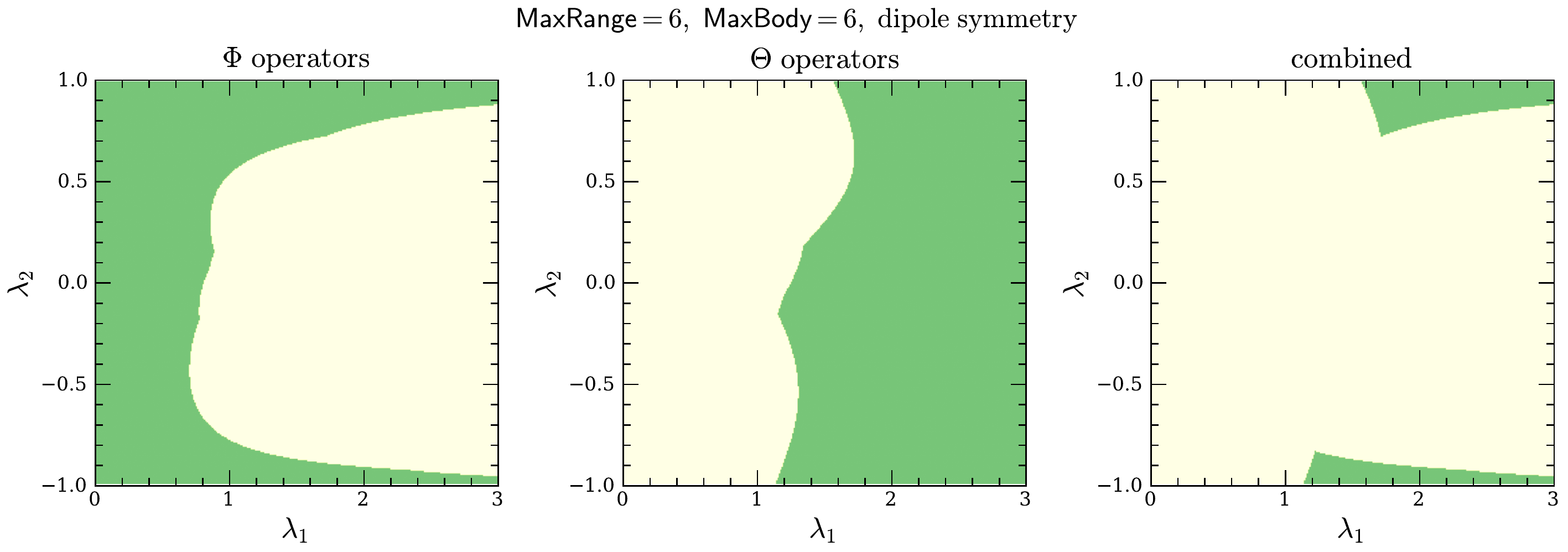}
	\caption{\label{fig:3panel_dipole_stab_fig} The same as in fig. \ref{fig:3panel_stab_fig}, but excluding $\cos(\cp_q)$ operators which have nonzero dipole moment.}
\end{figure*}

For the choice of $R^2(k_x,0)$ in \eqref{rchoice}, the analogue of fig. \ref{fig:3panel_stab_fig} in the presence of dipole symmetry is shown in fig. \ref{fig:3panel_dipole_stab_fig}. We see that imposing dipole symmetry leads to a slight increase in the size of the stability region, mostly in the region where $\l_2$ is close to $1$. 

Finally, note that the analysis for the case with dipole conservation but without translation symmetry is the same as the case with translation symmetry but without dipole symmetry, just with $R^2(k_x,0) \lra R^{-2}(k_x,0)$ (as translations act as a dipole symmetry on the $\t$ fields). 
%Therefore in the former case, a stable region can be obtained by taking $R^2(k_x,0)$ to be the inverse of the choice in \eqref{rchoice}. 

\ss{Stability in the presence of large marginal deformations} 

We found above that operators preserving the subsystem symmetries are always either marginal or irrelevant. For those operators which are marginal, we can then ask whether or not our conclusions about stability change if such operators are explicitly included in the fixed point action. 

\begin{widetext}

Focusing on operators built out of $\phi$, we found above that $\cos(\De_x^n\De_y^m\phi)$ is marginal provided that $\min(n,m)=1$. We may therefore consider adding these terms to the action and then expanding the cosines to quadratic order, yielding an action of the form 
\bea \label{marginals} S & = \frac{R^2}\twp \int d\tau \sum_i \Big( \frac1{2\vs} (\p_\tau\phi)^2 + \vs (\De_x\De_y\phi)^2  + \vs \sum_{n=2}^\infty\( \a_{x,n}(\De_x^n \De_y\phi)^2 + \a_{y,n}(\De_x\De_y^n\phi)^2\)\Big),\eea 
where the $\a_{a,n}$ are dimensionless constants, whose magnitudes we generically we expect to be smaller for larger values of $n$ (since microscopically such terms correspond to $2^{n+1}$-body boson operators). Nevertheless, as the added terms are marginal, from a field theory perspective it makes sense to treat them on the same footing as the leading $(\De_x \De_y \phi)^2$ ring-exchange term. 

The terms in the second line of \eqref{marginals} modify the dispersion relation of $\phi$ to 
\bea \label{modified_disp} \ep_\bfk & = 4\vs|\sin(k_xa/2)\sin(k_ya/2)| \( 1 + \sum_{n=2}^\infty4^{n-1}\( \a_{x,n} \sin^{2n-2}(k_xa/2) + \a_{y,n} \sin^{2n-2}(k_ya/2)\)\)^{1/2}. \eea
Since the added operators preserve all subsystem symmetries, they do not change the fact that the dispersion vanishes along the coordinate axes in momentum space. We will assume that the $\a_{a,n}$ are generic enough such that the dispersion continues to vanish only along the coordinate axes, that the first derivatives $(\p_{k_x}\ep)(0,k_y)$ and $(\p_{k_y}\ep)(k_x,0)$ continue to be non-vanishing for all nonzero $k_y$ and $k_x$ respectively, and that $(\p_{k_x}\p_{k_y}\ep)(\bfzero) \neq 0$. With these assumptions the RG eigenvalues of operators will still be determined by integrals over momentum shells surrounding the coordinate axes, and can computed using the same analysis as that appearing below \eqref{scaling_integral}. To leading order in $\eta = \L a  \ll 1$ (with the low energy modes defined as having momentum satisfying $\ep_\bfk \leq \vs \eta^2$, as before), we find 
\bea \label{general_dcoscp} \De_{\cp_q} & = \frac{a}{\twp} \int_{k_0}^{\pi/a}  \( \frac{dk_x \, |q(k_x,0)|^2 }{R^2(k_x,0) [(\p_{k_ya}\ep)(k_x,0)]} + \frac{dk_y \, |q(0,k_y)|^2 }{R^2(0,k_y)[ (\p_{k_xa}\ep)(0,k_y)]} \),\eea 
where $k_0 = \L / [(\p_{k_xa}\p_{k_ya}\ep)(\bfzero)]$ (which may be sent to zero when computing the scaling dimensions of the dipole operators). As a sanity check, note that this more general expression reduces to \eqref{dcoscp} upon setting $\ep_\bfk = 4 |\sin(k_xa/2)\sin(k_ya/2)|$.

%Since in condensed matter settings it seems reasonable to expect the $\a_{a,n}$ to all be rather small, and since the factors of $\sin^{2n-2}(k_{x/y}a/2)$ which contribute to the derivatives of $\ep$ appearing in \eqref{general_dcoscp} are always smaller than the leading $\sin(k_{x/y}a/2)$ term,  
By tuning the $\a_{a,n}$ appropriately it may very well be possible to render all of the dipole operators irrelevant, thereby producing an EBL phase stabilized by $(b\geq8)$-body ring-exchange terms. We leave a more detailed investigation of this possibility to the future.

\end{widetext}

\ss{Stability in the continuum limit}

All of the analysis in this paper has been concerned with the thermodynamic limit, where the lattice spacing is kept finite as the system size is sent to infinity. While this is the limit relevant to doing condensed matter physics, other types of limits, particularly continuum limits in which the lattice spacing is sent to zero, can be more natural in field theory settings. 

An interesting aspect of the EBL is that the physics is sensitive to the type of limit which is taken (see e.g. \cite{seiberg2003exotic,gorantla2021low}). In particular, in the continuum limit taken in Ref. \cite{gorantla2021low}, it was shown that {\it all} operators violating the particle and vortex subsystem symmetries are infinitely irrelevant, in the sense of having ultralocal correlation functions \cite{gorantla2021low}. Note that this includes the dipolar operators that we found to be responsible for destabilizing the $R^2 = {\rm const.}$ fixed point in our calculations. This means that conclusions about stability in the EBL depend crucially on the type of limit which is taken. It would be interesting to understand at a deeper level why this is so.

\section{Generalization to 3d}  \label{sec:3d} 

Everything we have discussed so far admits a straightforward generalization to bosons hopping on a cubic lattice in 3+1D, where we may consider a model whose kinetic term is dominated by a {\it cube-}exchange term
, as studied in refs \cite{you2020emergent,seiberg2020exotic,gorantla2020more}. 
The appropriate analogue of the free action \eqref{uvphiaction} is 
\be S = \frac{R^2}{\twp} \int d\tau \sum_i \( \frac1{2\vs}(\p_\tau \phi)^2 + \vs(1-\cos(\De_x\De_y\De_z\phi) )\),\ee 
where $\phi$ is again a field which keeps track of the phase of the UV bosons, with the boson density being written in terms of a dual field $\t$ as $n = \ob n + \frac1{\twp a^3} \De_x\De_y\De_z \t$. 
%This theory was aptly dubbed the ``XY-cube model'' in \cite{seiberg2020exotic,gorantla2020more}.  
The action of translation symmetry on $\t$ is, in analogy to \eqref{symm_action}, 
\be \label{3d_trans_action} T_\bfmu : \, \t(\bfr) \mt \t(\bfr+\bfmu) + \twp \ob n \( (x+\mu_x)(y+\mu_y)(z+\mu_z) - xyz\).\ee 
 Again as in the 2+1D case, this model as written has an infinite-dimensional group of linear subsystem symmetries, with the boson number being separately conserved along every line parallel to one of the coordinate axes. Again as in the 2+1D case, part of our task is to determine whether or not terms which break this gigantic symmetry group are relevant (in the technical sense). 

If we as before generalize to let $R^2$ be a function of $\bfk$ and work in the phase where the cube-exchange term dominates so that the cosine may be Taylor expanded, we obtain the Gaussian action  
\bea \label{general_gaussian_action_3d} S_0 = \int_{\bfk,\o}   \frac{R^2(\bfk)}{\fpi \vs} \( \o^2   + \ep_\bfk^2 \)|\phi(\o,\bfk)|^2, \eea 
where we have written $\int_{\bfk,\o} = \int \frac{d^2k}{(\twp/a)^3} \frac{d\o}{\twp}$, and where the dispersion is now
\be \label{disp_3d} \ep_\bfk = 8 \vs|\sin(k_xa/2) \sin(k_ya/2)\sin(k_za/2)|.\ee 
%The version where $R^2(\bfk)$ is independent of momentum has been studied in \cite{you2020emergent,seiberg2020exotic,gorantla2020more}. 

The dispersion \eqref{disp_3d} vanishes along the codimension-1 surface in momentum space spanned by the planes where one of $k_x,k_y,k_z$ vanishes. Given the analysis of the preceding sections it should be clear how to set up RG: scaling dimensions are controlled by the function $R^2(k_x,k_y,0)$ (assuming cubic symmetry), and the RG proceeds by integrating out shells with momentum satisfying 
\be \label{momentumshell_3d} (\eta')^3 < \ep_\bfk /\vs < \eta^3,\ee 
with $\eta=\L a\ll1$ and $\eta' = \eta(1-dt)$ as before. 

%The codimension-1 gapless surfaces in momentum space guarantee that the fluctuations in the model are quasi one-dimensional; as in the EBL this is enough to ensure that the theory can be analyzed through a Gaussian action depending on the phase field $\phi$ and its dual $\t$.  

%Therefore on a linear combination $\ct_q$ of $\t$ fields, translation acts as 
%\be T_\bfmu : \, \ct_q(\bfr) \mt \ct_q(\bfr+\bfmu) + \bfQ(q)\cdot\bfmu + \sum_{i\neq j\neq k}\bfD(q)_i \mu_j\mu_k + M(q)\mu_x\mu_y\mu_z,\ee 
%where $\bfQ(q)_i = \sum_a |\ep_{ijk}|q_i r^j_a r^k_a$ is the vortex quadrupole moment of $\ct_q$. 

We now define the scaling dimension of an operator $\cos(\mco)$ in terms of the associated fast mode propagator as (cf \eqref{sddef})
\be G_>(0,\bfzero) = 6 \De_\mco dt + O(dt^2).\ee 
The factor of 6 on the RHS is chosen so that correlation functions of $\mco$ are functions of spacetime distances to the power of $2\De_\mco$ (as can be shown along the lines of the calculations in appendix \ref{sec:app}).
%as discussed in appendix \ref{sec:app}. 
%This means the relevance of operators whose typical values are order 1 is determined by comparing $\De_\mco$ with 2, as in the two-dimensional case. 
The RG eigenvalue of a coupling $g$ associated with $\cos(\mco)$ is consequently
\be \label{3drgeig} y_g = d_\mco - 3\De_\mco,\ee 
with $d_\mco$ determined as before by requiring that when evaluated on typical field configurations, $\eta^{d_\mco}$ times the perturbation goes as $(\De_x\De_y\De_z\phi)^2 \sim \eta^6$ (for example if $\mco = \De_x\De_y \phi$, then $d_{\mco} = 6$). 

Evaluating $G_>$, we see that the scaling dimension of a general operator $\cos(\cp_q)$ is
\bea \label{general_3ddim} \De_{\cp_q} & = \frac1{6dt \cdot 8\pi^2} \int_{dS_\L} \frac{dk_x\, dk_y\, dk_z}{ R^2(\bfk) \L^3} |q(\bfk)|^2  \\ 
& = \frac{1}{32\pi^2} \(\int_{-\pi/a}^{-\L} + \int_\L^{\pi/a}\) dk_y\, dk_z\, \\ & \qq \times  \frac{|q(0,k_y,k_z)|^2}{R^2(0,k_y,k_z)|\sin(k_ya/2)\sin(k_za/2)|} + \cdots,\eea 
where $\cdots$ is a stand-in for analogous integrals over $dk_x\ dk_z$ and $dk_x\ dk_y$. 
From \eqref{general_3ddim} we again see that $\De_{\cp_q}$ diverges as $\eta\ra0$ unless $q(\bfzero)=0$, i.e. unless $\cos(\cp_q)$ is neutral under the global $U(1)$ boson number conservation. However unlike the two-dimensional case, we see that $\De_{\cp_q}$ also diverges unless $q(\bfk)$ vanishes when any {\it two} of $k_x,k_y,k_z$ vanish. Any operator with finite scaling dimension must therefore be neutral under all planar subsystem symmetries, i.e. must separately conserve the number of bosons in each lattice plane. Finally, we see that if $q(\bfk)=0$ whenever any {\it one} of $k_{x,y,z}=0$---i.e. if ${\cos(\cp_q)}$ is neutral under the linear subsystem symmetries---we have $\De_{\cp_q}=0$. As in the 2+1D case, the latter type of operators have $d_\mco \leq 0$, and as such they are always either marginal or irrelevant. 
All of the preceding statements apply equally well to $\cos(\ct_q)$ operators, with the only change being $R^2(\bfk)\lra R^{-2}(\bfk)$ in \eqref{general_3ddim}. 

%These facts mean that any while those which preserve the linear subsystem symmetries have vanishing scaling dimensions (and develop long-range order). Furthermore, from \eqref{general_3ddim} we see that the most relevant operators with non-infinite and nonzero scaling dimensions will be those which preserve the planar subsystem symmetries, and whose quadrupole moments are nonvanishing along only one direction, which wolog we may choose to be $\uvz$. Therefore the stability analysis reduces to that of a two-dimensional problem, with the relevant quadratic forms now being 
%\bea Q_\cp(q,q') & = \frac1{512\pi^2}\int_{-\pi/a}^{\pi/a} dk_x\, dk_y\, \frac{u(k_x,k_y)^*v(k_x,k_y)}{R^2(k_x,k_y,0) |\sin(k_xa/2)\sin(k_ya/2)|} \\ 
%Q_\ct(q,q') & = \frac1{512\pi^2}\int_{-\pi/a}^{\pi/a} dk_x\, dk_y\, \frac{R^2(k_x,k_y,0) u(k_x,k_y)^*v(k_x,k_y)}{ |\sin(k_xa/2)\sin(k_ya/2)|},\eea 
%where $u(k_x,k_y) = \sum_l e^{-ikl}u_l$, $u_l\in \rr$, with $l$ running over the sites of a two-dimensional lattice (and likewise for $q'(k)$). The XY-cube fixed point parameterized by $R^2(k_x,k_y,0)$ is stable provided that neither of the lattices $\mcl_\cp = \sqrt{Q_\vp}\zz^{2L}, \mcl_\vt = \sqrt{Q_\vt}\zz^{2L}$ contain a lattice point within a distance of 3 from the origin. 
%

From the above discussion, as long as we are only interested in operators with the potential to destabilize the Gaussian fixed point, we may without loss of generality restrict our attention to operators which 
are invariant under the planar subsystem symmetries and which have e.g. $q(0,k_y,k_z) = q(k_x,0,k_z)=0, q(k_x,k_y,0) \neq0$ (i.e., operators which have vanishing dipole moment, and have nonzero quadrupole moment oriented along $\uvz$). Any $\cp_q$ fitting the bill may be written as 
\bea \cp_q &= \sum_i q_i \big[\phi(\bfr_i) - \phi(\bfr_i+\uvx a) + \phi(\bfr_i + \uvx a + \uvy a) \\ & \qq - \phi(\bfr_i + \uvy a )\big],\eea 
where the $\{q_i\}$ are integers and 
where the sum is over the sites of a two-dimensional square lattice. In terms of the $\{q_i\}$, the scaling dimensions of interest may then be written as 
\bea \De_{\cp_q} & = \frac1{2\pi^2} \int_{-\pi/a}^{\pi/a} dk_x\, dk_y \frac{|\sin(k_xa/2)\sin(k_ya/2)| }{R^2(k_x,k_y,0)} \\ & \qq \times \sum_{i,j}q_iq_j \cos([\bfr_i-\bfr_j]\cdot \bfk),\\
\De_{\ct_q} & = \frac1{2\pi^2} \int_{-\pi/a}^{\pi/a} dk_x\, dk_y \, | \sin(k_xa/2)\sin(k_ya/2)|  \\ & \qq \times R^2(k_x,k_y,0) \sum_{i,j}q_iq_j \cos([\bfr_i-\bfr_j]\cdot \bfk).
\eea 
These operators all have $d_\mco = 6$, and hence from \eqref{3drgeig} their relevance is determined by comparing $\De_\mco$ with 2. 
Also note that by \eqref{3d_trans_action}, $\cos(\ct_q)$ is translation-invariant only if its (vortex) monopole, dipole, and quadrupole moments vanish. Therefore translation symmetry restricts to $\{ q_i\}$ such that $\sum_i q_i = 0$ in the second line above. 

\sss{Constant $R^2$}

Consider first the case where $R^2$ is independent of $\bfk$. The simplest potentially relevant cosine of the $\phi$ variables is $\cos(\De_x\De_y \phi)$, which has scaling dimension 
\be \De_{\De_x\De_y \phi} = \frac8{\pi^2 R^2},\ee 
therefore being irrelevant only when $R^2 < 4/\pi^2 \approx 0.4$. On the other hand, the simplest potentially relevant translation-invariant cosine of the $\t$ variables is e.g. $\cos(\De_x^2\De_y\t)$, with 
\be \De_{\De_x^2\De_y\t} = \frac{64R^2}{3\pi^2},\ee 
which is irrelevant only if $R^2>3\pi^2/32\approx 0.9$. Therefore the theory with constant $R^2$ is always unstable, as in the 2+1D case. However, if one imposes a global quadrupole symmetry on the $\phi$ fields, the simplest allowed $\cos(\cp_q)$ operator is then $\cp_q = \De_x^2 \De_y \phi$, which is irrelevant provided that $R^2<32/(3\pi^2) \approx 1.1$. There is then a small region $0.9 \lessapprox R^2 \lessapprox 1.1$ for which both this operator and $\cos(\De_x^2 \De_y \t)$ are irrelevant.  

\begin{figure*}
	\centering
	\includegraphics[width=.85\tw]{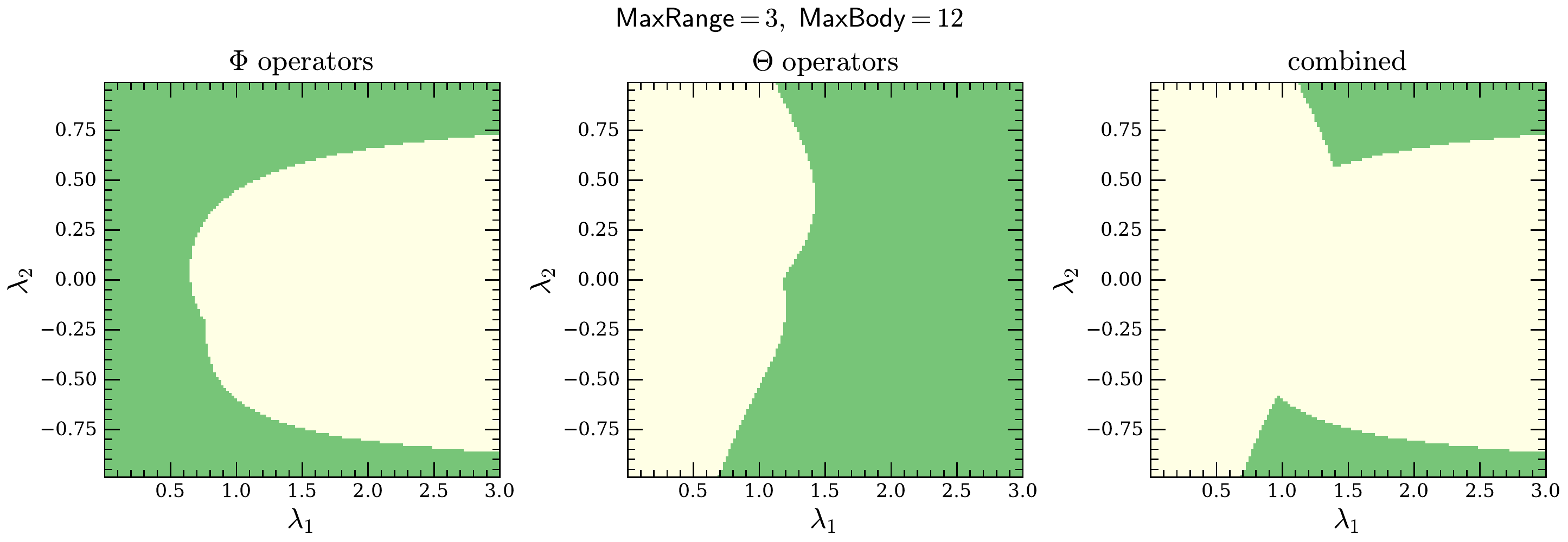}
	\caption{\label{fig:3panel_3d_stab_fig} Stability regions for the XY-cube model with the choice of $R^2(\bfk)$ given in \eqref{3drchoice}; translation symmetry is assumed.}
\end{figure*}

\sss{General $R^2(\bfk)$} 

For general choices of $R^2(\bfk)$ the story is similar to the 2d case, expect with slightly larger regions of stability for choices of $R^2(\bfk)$ analogous to that of \eqref{rchoice}. An example of the stability region in the $\l_1$-$\l_2$ plane for the choice 
\be  \label{3drchoice} R(\bfk) = \l_1(1+\l_2 \cos(a[k_x+k_y+k_z]))^2\ee 
is shown in fig. \ref{fig:3panel_3d_stab_fig}.

\section{Conclusion} \label{sec:conc}

In this note we have discussed a natural scheme for performing RG in the exciton Bose liquid and related models. We showed that although the simplest type of exciton Bose liquid is unstable within our RG scheme, a certain choice of marginal deformations can be made such that a stable phase is likely to be realizable. This last point was argued for on the basis of a simple numerical search, and it would be nice to obtain an analytic perspective on this issue, perhaps along the lines of that developed in \cite{plamadeala2014perfect}. 

More generally, this way of thinking about RG in models whose IR fixed points involve the appearance of a microscopic length scale may be useful in other contexts, e.g. in studying the 3+1D XY plaquette model \cite{seiberg2020exotic,gorantla2021low}. It would also be interesting to study RG flows in these models more generally, beyond just the rather elementary evaluation of RG eigenvalues performed here. 

%\ethan{need to email yizhi+zhen+pretko about their paper --- problem with their analysis of the $\p_x\p_y\p_z$ theory since they didn't consider cosines in the gauge-invariant variable $\t_-$}

\section*{Acknowledgments} 

I thank the University of Colorado Boulder and Marvin Qi for hospitality while this work was written up, and am grateful to Ho-Tat Lam and Shu-Heng Shao for discussions and detailed feedback on a first draft. I am supported by the Hertz Fellowship. 

\appendix 

\begin{widetext}

\section{Correlation functions} \label{sec:app} 

In this appendix we will compute a few correlation functions at the 2+1D EBL fixed point (the appropriate generalization to the 3+1D example of section \ref{sec:3d} is straightforward). See \cite{paramekanti2002ring,gorantla2021low} for a detailed analysis of related correlation functions in various different limits. 

We will focus on correlation functions of exponentials of $\cp_q$ operators defined as in \eqref{cpctdef}; correlators involving the $\t$ fields can be obtained by inverting $R^2(\bfk)$, as usual. 
Letting $S_\L$ denote the low-energy region in momentum space (where $\ep_\bfk \leq \vs\eta^2$), the two-point function of $e^{i\cp_q}$ is (working in units where $\vs=1$ for simplicity)
\bea -\ln \lan e^{i\cp_q(\bfr,\tau)} e^{-i\cp_q(\bfzero,0)}\ran & = \frac{a^2}\fpi \int_{S_\L} \frac{dk_x \, dk_y}{R^2(\bfk)}\frac{ |q(\bfk)|^2}{4 |\sin(k_xa/2)\sin(k_ya/2)|} \\ & \qq \times \( 1-(\cos(xk_x)\cos(yk_y) - \sin(xk_x)\sin(yk_y))e^{-4\tau |\sin(k_xa/2)\sin(k_ya/2)|}\) \eea 
where $\bfr = (x,y)$. Consider what needs to happen in order to cancel the logarithmic divergences that could arise when $k_x,k_y$ are small. 
%First, we see that as long as $\bfr,\tau$ do not both vanish, the RHS diverges unless $q(\bfzero)=0$. Therefore any operator with nonzero charge has correlation functions that are ultra-local in the thermodynamic limit. 
Sending $k_x\ra 0$ tells us that the RHS diverges unless either $y=0$ or $q(0,k_y)=0\,\, \forall\, \, k_y$, and likewise with $x\lra y$.
Therefore if $\bfr \neq0$ is parallel to $\uvx$ (to $\uvy$), the correlation function vanishes in the thermodynamic limit unless $\cp_q$ individually conserves the number of bosons along each column (each row) of the lattice. If $\bfr\neq0$ is not parallel to either of the coordinate axes, the correlation function vanishes unless $\cp_q$ respects both subsystem symmetries. In fact in this case, the correlator is asymptotically constant. Indeed, at all points in $S_\L$, we may always write $q(\bfk)$ as either $q(k_x,0)$ or $q(0,k_y)$, up to corrections vanishing as $\eta \ra 0$. As such, if $q(k_x,0) = q(0,k_y)=0$ for all $k_x,k_y$, the correlator is constant up to terms that vanish with $\eta \ra 0$. 

Consider now for simplicity the equal-time correlator, with $\tau=0$. From the comments above, the only interesting case is one where e.g. $\bfr= (0,y)$ and $q(0,k_y) = 0 \, \, \forall\, \, k_y$, but where $q(k_x,0)$ is nontrivial, and $y\neq0$. Then working up to terms that vanish as $\eta \ra 0$, we have  
\bea -\ln \lan e^{i\cp_q(\bfr,\tau)} e^{-i\cp_q(\bfzero,0)}\ran & = \frac{a^2}\twp \int_\L^{\pi/a} dk_x \int_0^{\eta^2/2a\sin(k_xa/2)} dk_y \frac{|q(k_x,0)|^2(1-\cos(yk_y))}{R^2(k_x,0)\sin(k_xa/2)k_ya}.\eea 
We now consider the limit of large spatial separation, where $y/a \gg 1/\eta^2$. In this regime we may perform the integral over $k_y$ to give 
\bea -\ln \lan e^{i\cp_q(\bfr,\tau)} e^{-i\cp_q(\bfzero,0)}\ran & = \frac{a}{\twp}\int_\L^{\pi/a} \frac{ dk_x \, |q(k_x,0)|^2}{R^2(k_x,0)\sin(k_xa/2)} \ln(y\eta^2/a),\eea
where we have only kept the leading piece in the $y/a\ra \infty$ limit. This means that for large $y$, we have 
\bea \lan e^{i\cp_q(0,y,0)} e^{-i\cp_q(\bfzero,0)}\ran \sim \frac1{|ya\L^2|^{2\De_{\cp_q}}},\eea 
where $\De_{\cp_q}$ is the scaling dimension as determined by taking $q(0,k_y) = 0$ in \eqref{dcoscp}. Note that as claimed in the main text, our definition of $\De_\mco$ ensures that the correlator is proportional to $|y|^{-2\De_{\cp_q}}$. 

On the other hand, consider the case where $\bfr=0,\tau\neq0$. We then have 
\bea -\ln \lan e^{i\cp_q(\bfzero,\tau)} e^{-i\cp_q(\bfzero,0)}\ran & = \frac{a^2}\fpi \int_{S_\L} \frac{dk_x \, dk_y}{R^2(\bfk)}\frac{ |q(\bfk)|^2(1-e^{-4\tau |\sin(k_xa/2)\sin(k_ya/2)|})}{4 |\sin(k_xa/2)\sin(k_ya/2)|} \\ 
& = \frac{a}{2\pi} \int_\L^{\pi/a} dk_x \int_0^{\eta^2/2a\sin(k_xa/2)} dk_y \frac{|q(k_x,0)|^2 (1-e^{-2\tau k_ya \sin(k_xa/2)})}{\sin(k_xa/2) k_y}  + (k_x\lra k_y).\\
 \eea 
 In the large $\tau$ limit where $\tau \gg 1/\eta^2$ (recall that we are in units where $\vs=1$) we may do the integrals whose upper limits go as $\eta^2$, and one can check that we obtain  
 \bea \lan e^{i\cp_q(\bfzero,\tau)} e^{-i\cp_q(\bfzero,0)}\ran \sim \frac1{|\tau\L^2|^{2\De_{\cp_q}}},\eea 
 where $\De_{\cp_q}$ is again as in \eqref{dcoscp}. The correlation functions when both $\tau$ and $\bfr$ are nonzero are obtained in a similar way. 

\end{widetext}

\bibliography{ebls}

\end{document}

%% file: ebls_preamble.tex
\usepackage{graphicx}
\usepackage[nointegrals]{wasysym} % for "\varhexagon" macro
\usepackage[export]{adjustbox}
\usepackage{amsmath,amsfonts,amssymb,latexsym}
\usepackage{hhline}
\usepackage{bm}
\usepackage{verbatim}
\usepackage{enumitem}
\usepackage{mathrsfs}
\usepackage{slashed}
\usepackage{empheq}

\newcommand{\p}{\partial}

\newcommand{\lan}{\langle}
\newcommand{\ran}{\rangle}

\newcommand{\da}{{\dagger}}

\newcommand{\ob}[1]{\mkern 1.5mu\overline{\mkern-1.5mu#1\mkern-1.5mu}\mkern 1.5mu}

\newcommand{\ra}{\rightarrow}
\newcommand{\lra}{\leftrightarrow}

\newcommand{\wt}{\widetilde}

\newcommand{\uva}{{\mathbf{\hat a}}}

\newcommand{\uvx}{{\mathbf{\hat x}}}
\newcommand{\uvy}{{\mathbf{\hat y}}}
\newcommand{\uvz}{{\mathbf{\hat z}}}

\newcommand{\bfzero}{{\mathbf{0}}}

\renewcommand{\(}{\left(}
\renewcommand{\)}{\right)}

\newcommand{\mt}{\mapsto}
\newcommand{\xra}{\xrightarrow}

\newcommand{\twp}{{2\pi}}
\newcommand{\fpi}{{4\pi}}

\newcommand{\D}{\nabla}

\newcommand\bpm            {\begin{pmatrix}}
	\newcommand\epm           {\end{pmatrix}}

\def\app#1#2{%
	\mathrel{%
		\setbox0=\hbox{$#1\sim$}%
		\setbox2=\hbox{%
			\rlap{\hbox{$#1\propto$}}%
			\lower1.1\ht0\box0%
		}%
		\raise0.25\ht2\box2%
	}%
}

\newcommand{\tw}{\textwidth}

\newcommand{\cp}{\Phi}
\newcommand{\ct}{\Theta}
\newcommand{\vs}{\varsigma}

\newcommand{\ope}\odot

\usepackage{manfnt}

\newcommand{\bi}{\begin{itemize}}
\newcommand{\ei}{\end{itemize}}

\usepackage{amsthm}

\newtheorem{theorem}{Theorem}

\theoremstyle{definition}
\newtheorem{definition}{Definition}
\theoremstyle{definition}

\newcommand\bd            {\begin{definition}}
\newcommand\ed            {\end{definition}}
\newcommand\bt            {\begin{theorem}}
\newcommand\et            {\end{theorem}}
\newcommand\be            {\begin{equation}}
\newcommand\ee            {\end{equation}}
\newcommand\ba            {\begin{aligned}}
\newcommand\ea            {\end{aligned}}
\newcommand\bea{\begin{equation}\begin{aligned}}
\newcommand\eea{\end{aligned}\end{equation}}

\usepackage{subcaption}

 \usepackage{hyperref} 
 \hypersetup{final}
 \hypersetup{colorlinks, citecolor=red, linkcolor=red, urlcolor=red} 
 
 \newcommand{\sss}{\subsubsection}
 \renewcommand{\ss}{\subsection}
 
 \renewcommand{\a}{\alpha}
 \renewcommand{\b}{\beta}
 \renewcommand{\d}{\delta}
 \newcommand{\De}{\Delta}
 
 \newcommand{\G}{\Gamma}

 \newcommand{\ep}{\varepsilon} %for historical reasons 
 \renewcommand{\l}{\lambda}
 \renewcommand{\L}{\Lambda}
 \renewcommand{\t}{\theta}
 
 \renewcommand{\o}{\omega}

 \newcommand{\bfmu}{{\boldsymbol{\mu}}}
 \newcommand{\bfnu}{{\boldsymbol{\nu}}}

 \newcommand{\bfk}{\mathbf{k}}
 \newcommand{\bfr}{\mathbf{r}}

 \newcommand{\nn}{\mathbb{N}}

 \newcommand{\rr}{\mathbb{R}}
 \newcommand{\qq}{\qquad}

 \newcommand{\zz}{\mathbb{Z}}

 \newcommand{\mck}{\mathcal{K}}

 \newcommand{\mco}{\mathcal{O}}
 \newcommand{\mcu}{\mathcal{U}}

 \usepackage[mathscr]{eucal} %this line gives mathscript that is less busy and closer to mathcal 

\usepackage{braket}

%% file: ebls_arxiv_version.bbl
\begin{thebibliography}{23}
\expandafter\ifx\csname natexlab\endcsname\relax\def\natexlab#1{#1}\fi
\expandafter\ifx\csname bibnamefont\endcsname\relax
  \def\bibnamefont#1{#1}\fi
\expandafter\ifx\csname bibfnamefont\endcsname\relax
  \def\bibfnamefont#1{#1}\fi
\expandafter\ifx\csname citenamefont\endcsname\relax
  \def\citenamefont#1{#1}\fi
\expandafter\ifx\csname url\endcsname\relax
  \def\url#1{\texttt{#1}}\fi
\expandafter\ifx\csname urlprefix\endcsname\relax\def\urlprefix{URL }\fi
\providecommand{\bibinfo}[2]{#2}
\providecommand{\eprint}[2][]{\url{#2}}

\bibitem[{\citenamefont{Paramekanti et~al.}(2002)\citenamefont{Paramekanti,
  Balents, and Fisher}}]{paramekanti2002ring}
\bibinfo{author}{\bibfnamefont{A.}~\bibnamefont{Paramekanti}},
  \bibinfo{author}{\bibfnamefont{L.}~\bibnamefont{Balents}}, \bibnamefont{and}
  \bibinfo{author}{\bibfnamefont{M.~P.} \bibnamefont{Fisher}},
  \bibinfo{journal}{Physical Review B} \textbf{\bibinfo{volume}{66}},
  \bibinfo{pages}{054526} (\bibinfo{year}{2002}).

\bibitem[{\citenamefont{You et~al.}(2020{\natexlab{a}})\citenamefont{You, Bibo,
  Pollmann, and Hughes}}]{you2020fracton}
\bibinfo{author}{\bibfnamefont{Y.}~\bibnamefont{You}},
  \bibinfo{author}{\bibfnamefont{J.}~\bibnamefont{Bibo}},
  \bibinfo{author}{\bibfnamefont{F.}~\bibnamefont{Pollmann}}, \bibnamefont{and}
  \bibinfo{author}{\bibfnamefont{T.~L.} \bibnamefont{Hughes}},
  \bibinfo{journal}{arXiv preprint arXiv:2008.01746}
  (\bibinfo{year}{2020}{\natexlab{a}}).

\bibitem[{\citenamefont{You et~al.}(2020{\natexlab{b}})\citenamefont{You, Bi,
  and Pretko}}]{you2020emergent}
\bibinfo{author}{\bibfnamefont{Y.}~\bibnamefont{You}},
  \bibinfo{author}{\bibfnamefont{Z.}~\bibnamefont{Bi}}, \bibnamefont{and}
  \bibinfo{author}{\bibfnamefont{M.}~\bibnamefont{Pretko}},
  \bibinfo{journal}{Physical Review Research} \textbf{\bibinfo{volume}{2}},
  \bibinfo{pages}{013162} (\bibinfo{year}{2020}{\natexlab{b}}).

\bibitem[{\citenamefont{Gorantla
  et~al.}(2021{\natexlab{a}})\citenamefont{Gorantla, Lam, Seiberg, and
  Shao}}]{gorantla2021low}
\bibinfo{author}{\bibfnamefont{P.}~\bibnamefont{Gorantla}},
  \bibinfo{author}{\bibfnamefont{H.~T.} \bibnamefont{Lam}},
  \bibinfo{author}{\bibfnamefont{N.}~\bibnamefont{Seiberg}}, \bibnamefont{and}
  \bibinfo{author}{\bibfnamefont{S.-H.} \bibnamefont{Shao}},
  \bibinfo{journal}{arXiv preprint arXiv:2108.00020}
  (\bibinfo{year}{2021}{\natexlab{a}}).

\bibitem[{\citenamefont{Seiberg and
  Shao}(2020{\natexlab{a}})}]{seiberg2003exotic}
\bibinfo{author}{\bibfnamefont{N.}~\bibnamefont{Seiberg}} \bibnamefont{and}
  \bibinfo{author}{\bibfnamefont{S.}~\bibnamefont{Shao}},
  \bibinfo{journal}{arXiv preprint arXiv:2003.10466}
  (\bibinfo{year}{2020}{\natexlab{a}}).

\bibitem[{\citenamefont{Tay et~al.}(2011)\citenamefont{Tay, Motrunich
  et~al.}}]{tay2011possible}
\bibinfo{author}{\bibfnamefont{T.}~\bibnamefont{Tay}},
  \bibinfo{author}{\bibfnamefont{O.~I.} \bibnamefont{Motrunich}},
  \bibnamefont{et~al.}, \bibinfo{journal}{Physical Review B}
  \textbf{\bibinfo{volume}{83}}, \bibinfo{pages}{205107}
  (\bibinfo{year}{2011}).

\bibitem[{\citenamefont{Kadanoff}(1966)}]{kadanoff1966scaling}
\bibinfo{author}{\bibfnamefont{L.~P.} \bibnamefont{Kadanoff}},
  \bibinfo{journal}{Physics Physique Fizika} \textbf{\bibinfo{volume}{2}},
  \bibinfo{pages}{263} (\bibinfo{year}{1966}).

\bibitem[{\citenamefont{You and Moessner}(2021)}]{you2021fractonic}
\bibinfo{author}{\bibfnamefont{Y.}~\bibnamefont{You}} \bibnamefont{and}
  \bibinfo{author}{\bibfnamefont{R.}~\bibnamefont{Moessner}},
  \bibinfo{journal}{arXiv preprint arXiv:2106.07664}  (\bibinfo{year}{2021}).

\bibitem[{\citenamefont{You et~al.}(2021)\citenamefont{You, Bibo, Hughes, and
  Pollmann}}]{you2021bfractonic}
\bibinfo{author}{\bibfnamefont{Y.}~\bibnamefont{You}},
  \bibinfo{author}{\bibfnamefont{J.}~\bibnamefont{Bibo}},
  \bibinfo{author}{\bibfnamefont{T.~L.} \bibnamefont{Hughes}},
  \bibnamefont{and} \bibinfo{author}{\bibfnamefont{F.}~\bibnamefont{Pollmann}},
  \bibinfo{journal}{arXiv preprint arXiv:2101.01724}  (\bibinfo{year}{2021}).

\bibitem[{\citenamefont{Gorantla
  et~al.}(2021{\natexlab{b}})\citenamefont{Gorantla, Lam, Seiberg, and
  Shao}}]{gorantla2021modified}
\bibinfo{author}{\bibfnamefont{P.}~\bibnamefont{Gorantla}},
  \bibinfo{author}{\bibfnamefont{H.~T.} \bibnamefont{Lam}},
  \bibinfo{author}{\bibfnamefont{N.}~\bibnamefont{Seiberg}}, \bibnamefont{and}
  \bibinfo{author}{\bibfnamefont{S.-H.} \bibnamefont{Shao}},
  \bibinfo{journal}{arXiv preprint arXiv:2103.01257}
  (\bibinfo{year}{2021}{\natexlab{b}}).

\bibitem[{\citenamefont{Shankar}(1994)}]{shankar1994renormalization}
\bibinfo{author}{\bibfnamefont{R.}~\bibnamefont{Shankar}},
  \bibinfo{journal}{Reviews of Modern Physics} \textbf{\bibinfo{volume}{66}},
  \bibinfo{pages}{129} (\bibinfo{year}{1994}).

\bibitem[{\citenamefont{Lake et~al.}(2021)\citenamefont{Lake, Senthil, and
  Vishwanath}}]{lake2021bose}
\bibinfo{author}{\bibfnamefont{E.}~\bibnamefont{Lake}},
  \bibinfo{author}{\bibfnamefont{T.}~\bibnamefont{Senthil}}, \bibnamefont{and}
  \bibinfo{author}{\bibfnamefont{A.}~\bibnamefont{Vishwanath}},
  \bibinfo{journal}{Physical Review B} \textbf{\bibinfo{volume}{104}},
  \bibinfo{pages}{014517} (\bibinfo{year}{2021}).

\bibitem[{\citenamefont{Lake}(to appear)}]{lake2021fermi}
\bibinfo{author}{\bibfnamefont{E.}~\bibnamefont{Lake}} (\bibinfo{year}{to
  appear}).

\bibitem[{\citenamefont{Vishwanath et~al.}(2004)\citenamefont{Vishwanath,
  Balents, and Senthil}}]{vishwanath2004quantum}
\bibinfo{author}{\bibfnamefont{A.}~\bibnamefont{Vishwanath}},
  \bibinfo{author}{\bibfnamefont{L.}~\bibnamefont{Balents}}, \bibnamefont{and}
  \bibinfo{author}{\bibfnamefont{T.}~\bibnamefont{Senthil}},
  \bibinfo{journal}{Physical Review B} \textbf{\bibinfo{volume}{69}},
  \bibinfo{pages}{224416} (\bibinfo{year}{2004}).

\bibitem[{\citenamefont{Fradkin et~al.}(2004)\citenamefont{Fradkin, Huse,
  Moessner, Oganesyan, and Sondhi}}]{fradkin2004bipartite}
\bibinfo{author}{\bibfnamefont{E.}~\bibnamefont{Fradkin}},
  \bibinfo{author}{\bibfnamefont{D.~A.} \bibnamefont{Huse}},
  \bibinfo{author}{\bibfnamefont{R.}~\bibnamefont{Moessner}},
  \bibinfo{author}{\bibfnamefont{V.}~\bibnamefont{Oganesyan}},
  \bibnamefont{and} \bibinfo{author}{\bibfnamefont{S.~L.}
  \bibnamefont{Sondhi}}, \bibinfo{journal}{Physical Review B}
  \textbf{\bibinfo{volume}{69}}, \bibinfo{pages}{224415}
  (\bibinfo{year}{2004}).

\bibitem[{\citenamefont{Lam}(private communication)}]{hotat}
\bibinfo{author}{\bibfnamefont{H.-T.} \bibnamefont{Lam}}
  (\bibinfo{year}{private communication}).

\bibitem[{\citenamefont{Xu and Fisher}(2007)}]{xu2007bond}
\bibinfo{author}{\bibfnamefont{C.}~\bibnamefont{Xu}} \bibnamefont{and}
  \bibinfo{author}{\bibfnamefont{M.~P.} \bibnamefont{Fisher}},
  \bibinfo{journal}{Physical Review B} \textbf{\bibinfo{volume}{75}},
  \bibinfo{pages}{104428} (\bibinfo{year}{2007}).

\bibitem[{\citenamefont{Plamadeala et~al.}(2014)\citenamefont{Plamadeala,
  Mulligan, and Nayak}}]{plamadeala2014perfect}
\bibinfo{author}{\bibfnamefont{E.}~\bibnamefont{Plamadeala}},
  \bibinfo{author}{\bibfnamefont{M.}~\bibnamefont{Mulligan}}, \bibnamefont{and}
  \bibinfo{author}{\bibfnamefont{C.}~\bibnamefont{Nayak}},
  \bibinfo{journal}{Physical Review B} \textbf{\bibinfo{volume}{90}},
  \bibinfo{pages}{241101} (\bibinfo{year}{2014}).

\bibitem[{\citenamefont{Lake and Hermele}(2021)}]{lake2021subdimensional}
\bibinfo{author}{\bibfnamefont{E.}~\bibnamefont{Lake}} \bibnamefont{and}
  \bibinfo{author}{\bibfnamefont{M.}~\bibnamefont{Hermele}},
  \bibinfo{journal}{arXiv preprint arXiv:2107.09073}  (\bibinfo{year}{2021}).

\bibitem[{\citenamefont{Mukhopadhyay et~al.}(2001)\citenamefont{Mukhopadhyay,
  Kane, and Lubensky}}]{mukhopadhyay2001sliding}
\bibinfo{author}{\bibfnamefont{R.}~\bibnamefont{Mukhopadhyay}},
  \bibinfo{author}{\bibfnamefont{C.}~\bibnamefont{Kane}}, \bibnamefont{and}
  \bibinfo{author}{\bibfnamefont{T.}~\bibnamefont{Lubensky}},
  \bibinfo{journal}{Physical Review B} \textbf{\bibinfo{volume}{64}},
  \bibinfo{pages}{045120} (\bibinfo{year}{2001}).

\bibitem[{\citenamefont{Vishwanath and Carpentier}(2001)}]{vishwanath2001two}
\bibinfo{author}{\bibfnamefont{A.}~\bibnamefont{Vishwanath}} \bibnamefont{and}
  \bibinfo{author}{\bibfnamefont{D.}~\bibnamefont{Carpentier}},
  \bibinfo{journal}{Physical review letters} \textbf{\bibinfo{volume}{86}},
  \bibinfo{pages}{676} (\bibinfo{year}{2001}).

\bibitem[{\citenamefont{Seiberg and
  Shao}(2020{\natexlab{b}})}]{seiberg2020exotic}
\bibinfo{author}{\bibfnamefont{N.}~\bibnamefont{Seiberg}} \bibnamefont{and}
  \bibinfo{author}{\bibfnamefont{S.-H.} \bibnamefont{Shao}},
  \bibinfo{journal}{arXiv preprint arXiv:2004.00015}
  (\bibinfo{year}{2020}{\natexlab{b}}).

\bibitem[{\citenamefont{Gorantla et~al.}(2020)\citenamefont{Gorantla, Lam,
  Seiberg, and Shao}}]{gorantla2020more}
\bibinfo{author}{\bibfnamefont{P.}~\bibnamefont{Gorantla}},
  \bibinfo{author}{\bibfnamefont{H.~T.} \bibnamefont{Lam}},
  \bibinfo{author}{\bibfnamefont{N.}~\bibnamefont{Seiberg}}, \bibnamefont{and}
  \bibinfo{author}{\bibfnamefont{S.-H.} \bibnamefont{Shao}},
  \bibinfo{journal}{SciPost Phys} \textbf{\bibinfo{volume}{9}},
  \bibinfo{pages}{2007} (\bibinfo{year}{2020}).

\end{thebibliography}
